\documentclass[fleqn,usenatbib]{mnras}

\usepackage{newtxtext,newtxmath}

\usepackage[T1]{fontenc}

\DeclareRobustCommand{\VAN}[3]{#2}
\let\VANthebibliography\thebibliography
\def\thebibliography{\DeclareRobustCommand{\VAN}[3]{##3}\VANthebibliography}

\usepackage{graphicx}	

\usepackage{amsmath,mathrsfs,epstopdf,slashed,color}
\usepackage{tikz}
\usetikzlibrary{matrix}
\usetikzlibrary{positioning}
\usepackage{fancyhdr} 
\usepackage{lastpage} 
\usepackage{extramarks} 
\usepackage{lipsum} 
\usepackage{amsfonts}
\usepackage{latexsym}
\usepackage{xypic}
\usepackage{cancel,slashed}

\usepackage{graphicx}
\usepackage{placeins}
\usepackage{hyperref}
\usepackage{cite}
\usepackage{braket}
\usepackage{bm}
\usepackage{diagbox}
\usepackage{multirow}
\usepackage{subcaption}

\defcitealias{lvk-gwtc3-data-paper}{LVK}
\defcitealias{lvk-hubble}{LVK}
\defcitealias{lvk-NSBH-discovery-paper}{LVK}
\defcitealias{gw17-bns-data}{LIGO-Virgo}
\defcitealias{gw17-bns-hubble}{LIGO-Virgo}
\defcitealias{gwtc2.1-catalog}{LIGO-Virgo}
\defcitealias{gwtc3-bbh-population}{LVK}
\defcitealias{gwtc2-bbh-population}{LIGO-Virgo}
\newcommand\mycitep[1]{\citepalias[\citeyear{#1}]{#1}}
\newcommand\mycitet[1]{\citetalias{#1} (\citeyear{#1})}  

\title[Measuring $H_0$ with NSBH via GW]{Pure Gravitational Wave Estimation of Hubble's Constant  \\ using Neutron Star - Black Hole Mergers}

\author[L. Fung et.al.]{
Leo W.H. Fung,$^{1}$ \thanks{leowhfung@gmail.com}
Tom Broadhurst ,$^{2,3,4}$
George F. Smoot $^{1,3,5,6,7}$
\\
$^1$ Department of Physics and Jockey Club Institute for Advanced Study,
			Hong Kong University of Science and Technology, Hong Kong
	\\		
$^2$ {University of the Basque Country UPV/EHU, Department of Theoretical Physics, Bilbao, E-48080, Spain} \\
$^3$ {DIPC, Basque Country UPV/EHU, San Sebastian, E-48080, Spain}\\
$^4$ {Ikerbasque, Basque Foundation for Science, Bilbao, E-48011, Spain}
\\
$^5$ {Energetic Cosmos Laboratory, Nazarbayev University, Nursultan, Kazakhstan}
\\
$^6$ APC; AstroParticule et Cosmologie, Universit\'{e} de Paris, Paris Centre for Cosmological Physics,  CNRS/IN2P3\\ \, CEA/lrfu, 10, rue Alice Domon et Leonie Duquet,	75205 Paris CEDEX 13, France  {\it Emeritus}
\\
$^7$ {Physics Department \& LBNL, University of California at Berkeley CA 94720 {\it Emeritus}}
}

\date{Accepted XXX. Received YYY; in original form ZZZ}

\pubyear{2023}

\begin{document}
\label{firstpage}
\pagerange{\pageref{firstpage}--\pageref{lastpage}}

\maketitle

\begin{abstract}
Here we show how $H_0$ can be derived purely from the gravitational waves (GW) of neutron star-black hole (NSBH) mergers. This new method provides an estimate of $H_0$ spanning the redshift range, $z<0.25$ with current GW sensitivity and without the need for any afterglow detection. We utilise the inherently tight neutron star mass function together with the NSBH waveform amplitude and frequency to estimate distance and redshift respectively, thereby obtaining $H_0$ statistically. Our first estimate is $H_0 = 86^{+55}_{-46}$ km s$^{-1}$ Mpc$^{-1}$ for the secure NSBH events GW190426 and GW200115. We forecast that soon, with 10 more such NSBH events we can reach competitive precision of $\delta H_0/H_0 \lesssim 20\%$. 
\end{abstract}




\section{Introduction}
    Knowledge of the 
    expansion rate of the universe is essential for deriving the physical characteristics of extragalactic objects, including sizes and masses of galaxies and clusters and the age and luminosity of their stellar contents. 
    Increasingly precise standard candles and rulers have revealed the Hubble expansion has actually accelerated over the past few Gyrs, indicating a constant energy density with negative pressure governs the expansion rate today. As $H_0$ measurements have improved in precision, consistent claims that smaller values of $H_0$ have been inferred from distant cosmological measurements compared with local estimates of $H_0$ from the stellar calibrated standard candles \citep{pantheon-sne,sh0es-hubble, hubble-tension-review,hubble-hunter}. 
    A careful recalibration of local cephieds is now claimed to slightly relieve this $H_0$ tension when supernovae are classified according to metallicity, or from recalibration of evolved stars at the tip of red giant branch \citep{h0-trgb-2019,h0-trgb-2020} (and see \citet{h0-trgb-review} for a review).
  
    Alternatively, theoretical proposals to explain the Hubble Tension include a re-interpretation of the cosmic microwave background (CMB) data, invoking the modifications of recombination physics (for a review \citet{hubble-tension-solutions}), may allow the `standard' sound horizon scale length to be slightly smaller, by shifting the surface of last scattering to a little earlier in time, corresponding to a higher redshift allowing agreement with the local $H_0$ estimates. It has been shown this may be achieved by modifying recombination physics \citep{axi-higgs,axi-higgs-letter,axi-higgs-boltzmann,recombination-sound-horizon, cmb-t0-tension} that shifts the calibration, allowing room for reinterpreting the data with a somewhat higher $H_0$. 

    The seemingly irreconcilable differences among $H_0$ measurements obtained from different `standardiztable' calibration (candles and rulers) strengthen the need for measuring $H_0$ with independent methods with calibrations driven by unrelated physical mechanisms. 
    Gravitational wave (GW) observations provide exciting new opportunities for constraining such cosmological expansion. 
    Compared to conventional electromagnetic (EM) observations where flux falls as the square of the luminosity distance, the GW signal strength scales only linearly inverse with luminosity distance - as the detected strain amplitude depends on the square root of energy density per unit time. 
    As a result, the scaling of GW signal strength can be used to distinguishing models that involve modification of photon trajectories (consequently modifying the cosmic distance duality relation \citep{etherington-ratio,cosmic-distance-duality-relation-theory}), or other proposals that introduce strongly redshift dependent physics (for example \citet{late-phantom-de}).

	
	Aside from the unique scaling of signal strength, there are also some applications of GW that depend on physical assumptions lying outside gravitational wave data and consequently, such methods do not fully benefit from the new GW-based window.
    It has been proposed that spatial cross correlations between GW sources and galaxy surveys can provide a `siren' based angular scale that does not rely on individual (unknown) GW host galaxy identification. By using galaxy redshift surveys of suitable depth \citep{mukherjee-h0-angular,mukherjee-h0-angular-2,gw-hubble-cross-correlation-oguri} and all BBH events to date a value of $H_0=68^{+8}_{-6} \mathrm{km s}^{-1} \mathrm{Mpc}^{-1}$ has been estimated. 
    This rests on a statistical comparison of the ensemble of BBH sky location areas with the Dark Energy Survey (DES) optical galaxy survey as no unique host galaxy identification \mycitep{lvk-hubble} is yet possible. 
    Hence, this type of method that relying on the establishment of a standard ruler in general shares almost the same set of systematic problems as traditional standard ruler approaches, in particular the baryonic acoustic oscillations and as a result, the GW data does not represent a truly independent means for constraining $H_0$.
    For the consideration of Hubble tension, this would be unable to offer an independent check of the systematics.
    To overcome this limitation, we propose to rely on pure GW emission, as described below. 
    


    The long anticipated GW-based method  to constrain $H_0$ using binary neutron stars, pioneered by \citep{original-bns-hubble-Schutz-1986}, falls in the `standard candle' class and requires a prompt detection of nova emissison generated during the merger, to define the redshift.  This method has finally been realised with the first GW detection of a BNS event GW170817, for which  there is an optical redshift \mycitep{gw17-bns-data,gw17-bns-hubble}. Fortunate Fermi satellite detection of prompt gamma-ray emission from associated kilonova emission has localised the host galaxy, providing an optical spectroscopic redshift for the host. The recession velocity of the host galaxy is only $3000$ km s$^{-1}$ requiring a sizeable peculiar motion correction, at an estimated level of $\simeq 800$ km s$^{-1}$, resulting in $H_0= 70^{+12}_{-8.0} \mathrm{km\, s}^{-1} \mathrm{Mpc}^{-1}$. 
    The uncertainty on this estimate brackets the disputed range of $H_0$, and implies approximately 10 more such BNS events are required to address independently the current $H_0$ tension between SN-Ia and the CMB/Large Scale Structure (LSS) based estimates \citep{bns-hubble-Nissanke-analysis}.
    However, such coincident, multi-messenger observations are not anticipated to be available frequently, unless the angular resolution of GW observatories can be drastically improved (by $\sim 3$ orders of magnitude) to localized the position of the event, and indeed subsequent events classified as BNS were not detected at other wavelengths in prompt follow-up searches.

	These GW based methods for $H_0$ rely on an independent EM based redshift estimation, either individually in the case of BNS mergers or statistically with redshift surveys described above. 
    
    A pure GW based method for estimating $H_0$ has been proposed for the future, requiring much improved sensitivity, when the tidal deformation of the NS by the BH during a NSBH merger is detectable. This can provide in principle a GW based estimate of $H_0$ from the additional distance dependence on this physical deformation \citep{ns-love-hubble}. Here we propose another more practical `standard' strain pure GW method, that has apparently been overlooked, provided by the inherently narrow NS mass function established by radio, X-ray and optical observations in the Milky Way and local group, including binary neutron star pulsars. 
    
    We will show below that the GW radiation from neutron star-black hole (NSBH) binaries, rather than BNS events, is more suitable for the purpose of derive a reliable estimate of $H_0$, free from local peculiar motions that introduce model dependence for BNS estimates. We apply this method to the currently secure NSBH event GW200115 and other possible and proposed candidate NSBH events variously reported. We demonstrate that the existing instruments \citep{aligo-o3-instrument} already suffice for constraining $H_0$ within $\delta H_0/H_0 \lesssim 20\%$ simply by gathering $10$ years of measurements, even in the absence of significant sensitivity upgrades that may be anticipated. 
    We will emphasize throughout that NSBH events and to a lesser extent BNS events, can in combination provide competitive precision on $H_0$ that is purely derived from the GW radiation alone, without the need for a spectroscopic redshift.  
    Such additional independent information would of course provide further improvement, so that this combined general NS based estimate of $H_0$ we can anticipate will prove useful in practice.

	
	Our paper is organized as follows: in Section~\ref{sect: framework}, we sketch out the basis of the method, and then we describe our choice of datasets for GW events and local X-ray binaries in Section~\ref{sect: data}. We present a rigorous Bayesian hierarchical analysis applied to the data in Section~\ref{sect: data-analysis} and preset our analysis of the first reported GW events compatible with being NSBH binaries, focusing in particular on the secure NSBH detection claimed to date, GW200115 \mycitep{lvk-NSBH-discovery-paper} for a first estimate of $H_0$. We the make a forecast for the precision attainable for $H_0$, and the evaluation of associated systematics in Section~\ref{sect: forecast}. We finally conclude in Section~\ref{sect: conclusion}.

	\section{Intuitive Overview of the Method} \label{sect: framework}
 
    The widely applied standard candle measurement of $H_0$ compares independently determined luminosity distances, $d_L$, with spectroscopic redshifts, $z$, for type-Ia supernovae where luminosity is calibrated with local Cepheids. Here we emphasises that a pure GW based measure of $H_0$ is now feasible without the need for a follow-up spectroscopic redshift. Owing to stellar stability constraints, the mass range of neutron stars is fairly tight about the Chandrasekhar mass.
    Hence the lower mass binary member (neutron star) modulation of the NSBH waveform is predictable, so the degeneracy between luminosity distance and redshift of NSBH is relatively small.
    Consequently GW detections alone provide useful measurements of both $d_L$ and $z$ and hence a statistical measure of $H_0$, as we outline below. 
    This method we apply to GW waveforms identified as NSBH binaries, where the mass ratio can be defined from the waveform modulation and the presence of the relatively massive BH extends the detection horizon to $z \lesssim 0.25$, large enough for statistical measure of $H_0$.

	\subsection{Observables} 
	
	For our purposes we require only the time varying GW strain amplitude and phase registered by the laser interferometers, rather than relying on the template fitted waveform derived parameters reported for the GW detections. We emphasize here non-parametric (independent of source model) summary statistics that can be used to approximately describe the waveform below with post-Newtonian (PN) calculations that allow us to describe the physical quantities of the GW source with the summary statistics. Below, we will use the subscript $ \,_{\rm obs}$ to denote these non-parametric statistics.
	
	The amplitude of the strain $h_{\rm obs}$ according to the post-Newtonian (PN) calculations \citep{pn-calculation,Ajith-2007} for compact binary sources is given by the following combination of physical parameters:
	
	\begin{equation}
	    h_{\rm obs} \propto \frac{((1+z)\mathcal{M}_c)^{5/6}}{d_L}
	\end{equation}
	The $(1+z)$ factor can be understood as follows: as $h_{\rm obs}$ scales with $\sqrt{\rm Energy}$, the rate of radiation energy received by the observer is time dilated by the cosmological redshift. Also, the inverse square law $1/d_L^2$ diffusion of energy (flux) implies an inverse dependence of $h_{\rm obs}$ on luminosity distance $d_L$.  Hence, to determine $d_L$, we therefore need another observable, ideally independent to $h_{\rm obs}$, to pin down $(1+z)\mathcal{M}_c$. If we take the fourier transform of the time series of strain over different time windows (for example, Q-transform), the time evolution of the corresponding frequency spectrum can also be determined. 
	\footnote{The situation is more complicated when the signal is contaminated by noise, where parametric models of the GW waveform are used as matched filters  for deriving the waveform evolution. We will ignore this point in this section for illustration.}
	The first quantity is the redshifted chirp mass $(1+z)\mathcal{M}_c$. The chirp mass is defined as:
	\begin{equation} \label{eq: chirp-mass}
	\mathcal{M}_c \equiv \frac{(m_1m_2)^{3/5}}{(m_1 + m_2)^{1/5}} = \frac{1/q^{3/5}}{(1+1/q)^{1/5}} m_2,
	\end{equation}
	where we have defined the mass ratio $q=m_2/m_1$, and  implicitly by definition $m_2 < m_1$ for sorting the masses in the binary system. 
	This quantity, upon redshifted, is indeed imprinted on the spectral moment of the 0.5PN chirping spectrum $f_{\rm obs}(t)$: 
    (note however, that solving the spectral evolution must be done by solving the full differential equation, and the order-by-order approach as outlined below only make sense under some strict conditions. 
    \footnote{Strictly speaking, the spectral moments cannot be separated in this way. However, by using multi-scale perturbation methods (for example, Poincare-Lindstedt), the order-by-order separation can be a good approximation in a narrow time window, where the window time scale is defined by the leading coefficients of each spectral moment. 
    }
    )
	\begin{equation} \label{eq: redshifted-chirp-mass}
	    f_{\rm obs}^{-11/5} (\frac{df_{\rm obs}}{dt})^{3/5} \sim (1+z)\mathcal{M}_c.
	\end{equation}
	The redshift effect is easy to understand: the frequency of GW in the source frame is redshifted to $(1+z)^{-1}f_{\rm obs}$ and thus the time derivative $(1+z)^{+1}t_{\rm obs}$. 
	If we can pin down $m_2$ and $q$, it would then be possible to estimate $z$. Indeed, considering NSBH system, the neutron star mass $m_2$ can be pinned down by Milky Way observations as $m_2 = 1.4 \pm 0.1 M_\odot$ (see Section~\ref{sect: data} for the evidence of this number). 
	
	On an order-by-order basis, it is also possible to determine the mass ratio $q$ by another combination of spectral evolution:
	\begin{equation}
	    f_{\rm obs}^{-13/3} \frac{df_{\rm obs}}{dt} \sim \frac{1/q}{(1+1/q)^{2}},
	\end{equation}
	where all the redshift dependencies are cancelled by the multiplicative scaling term that we have neglected here. \citep{pn-calculation} 
	\footnote{Precisely, the spectral evolution $df_{\rm obs}/dt$ couples with the polynomial in $f_{\rm obs}$. As a result, the combination of moments of the spectrum are not orthogonal to each other. The full PN calculation for relating the observables to the chirp spectrum can be found in the pioneering work \citep{pn-calculation} (in particular Eq.1.3 therein) and for accurate estimation the spectrum must be predicted numerically using ODE solvers. }
	Therefore, this observable provides a measurement on $q$.
	
	\begin{figure*}
	    \centering
	    \includegraphics[width=\textwidth]{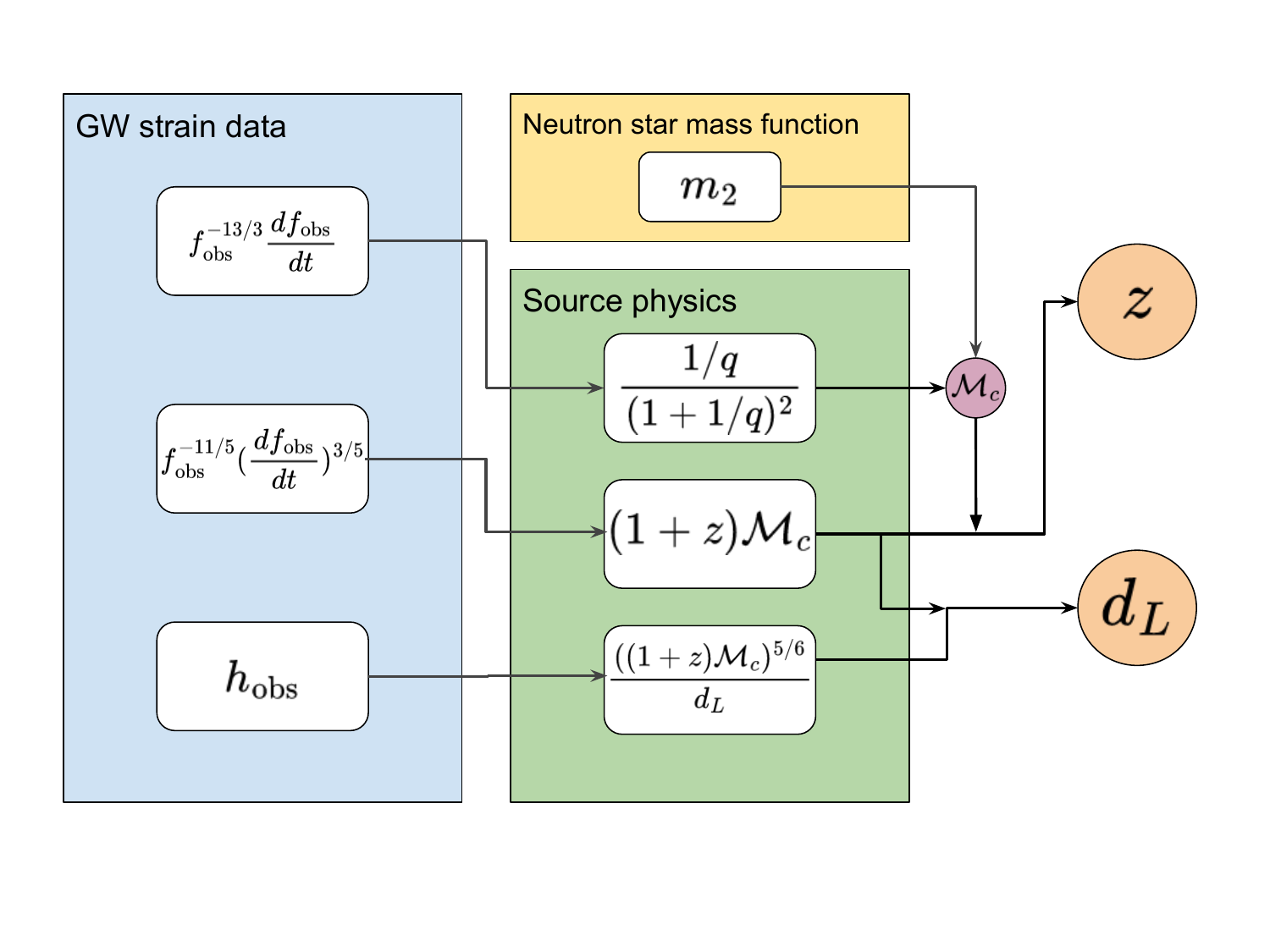}
	    \caption{Summary of the relations between observables and the physical quantities of the GW source, and the associated relation to quantities of cosmological interest.}
	    \label{fig: framework-summary}
	\end{figure*}
	
	We summarize the relations between the observables and physical source quantities in Fig.~\ref{fig: framework-summary}. Clearly, the observed quantities provide largely independent measurements of $z$ and $d_L$ respectively. The $(z_i, d_{L,i})$ pairs from each observed NSBH event $i$ can therefore be used to determine the Hubble constant $H_0$. In the big bang cosmology, we have:
    \begin{equation}
        \frac{d_L}{(1+z)} = \frac{c}{H_0} \int^z_0 \frac{dz'}{E(z')}.
    \end{equation}
    In the subsequent numerical analysis, we will assume $\Omega_m = 0.3$, so that $E(z) = \sqrt{0.3(1+z)^3 + 0.7}$, as the current data quality does not allow strong constrain on $\Omega_m$ together with $H_0$.

\subsection{Comparison with Binary Neutron Star events}

    Here we compare and contrast the ability of BNS and NSBH events for estimating $H_0$ by the method proposed here, including BNS events that do not have an independent spectroscopic redshift estimate because no host galaxy nor EM emission is identified.
    The absence of observed EM counterparts likely comprises the majority of such NS related GW events given the follow-up success to date, with EM emission found for just one very local BNS event, at only $40$ Mpc and the absence of EM detections for the other BNS and NSBH candidates.
	
	One may argue BNS systems are more suitable for the purpose of establishing an independent estimate on $z$ and thus $H_0$. 
    This is seemingly evident from Eq.\ref{eq: chirp-mass}: if both $m_{s,1}$ and $m_{s,2}$ are pinned down, $\mathcal{M}_{c}$ are known in a priori. 
    There are though two advantages for NSBH events compared to BNS in terms of realising our $H_0$ method. Firstly, the detection horizon for NSBH observations is much larger than is BNS observations. As shown in \citep{Ajith-2007}, the signal-to-noise ratio (SNR) of GW events scale as $\rho(m_{s,1}, m_{s,2}, z) \sim \frac{((1+z)\mathcal{M}_{c})^{5/6}}{d_L(z)}$. 
	For BNS with mass ratio $q \approx 1$, $\mathcal{M}_{c} \approx 0.87 m_2$; whereas the BH component serves as an efficient amplifier, boosting the detectable SNR of the companion NS: on average, the BH mass can be as high as $m_{s,1}\approx 20 M_\odot$, corresponding to an amplification factor of $1/q\approx 13$. Thus $\mathcal{M}_{c} \approx 2.75 m_2$, thereby extending the detection horizon for NSBH events to $d_L^{(\rm NSBH)}/d_L^{\rm (BNS)} \approx 4$, four times times larger than BNS events. 
	
	The larger horizon depth for NSBH event detection relative to BNS events can be translated into an improvement in $H_0$ via a simple Fisher information argument. Roughly the likelihood for $H_0$ follows a Gaussian $\mathcal{G}( d_L ; \hat{d}_L(z;H_0), \sigma_d )$, where the expected luminosity distance $\hat{d}_L$ at redshift $z$ is parameterized by $H_0$, subjected to distance measurement uncertainty $\sigma_d$. The Cramor-Rao bound states the propagation of uncertainty to $H_0$ is roughly (with the angled bracket denotes `taking expectation' from the distribution):
	\begin{equation} \label{eq: h0-uncertainty-fisher}
	\begin{split}
	    \delta{H_0} \geq \left\langle \frac{\partial \log \mathcal{G}}{\partial H_0}^2 \right\rangle^{-1/2}_{\sim d_L} &= \left\langle (\frac{\partial \log \mathcal{G}}{\partial \hat{d}_L}\frac{\partial \hat{d}_L}{\partial H_0})^2 \right\rangle^{-1/2}_{\sim d_L} \\
	    &\sim \sigma_d H_0 \frac{1}{\hat{d}_L},
	\end{split}
	\end{equation}
    where $\delta H_0$ denotes the standard deviation of the $H_0$ posterior.
	Therefore, $\delta H_0$ scales (at leading order) as $1/\hat{d}_L$: the enlarged horizon allows more amplification of the difference between a static universe and an expanding universe, hence helping to shrink the uncertainty in $H_0$, provided that the distance measurement uncertainty $\sigma_d$ scales slower than $\hat{d}_L$.

    The enlarged detection horizon is also beneficial in terms of the precision in pinning down the cosmological redshift. 
    In practice, the measured redshift is contributed by both the cosmological expansion and the local peculiar motion.
    When the fractional redshift error is assessed, a low detection horizon would be translated to a larger fractional redshift uncertainty. 
    This effect can be mitigated by making the peculiar motion-induced redshift contributes only a tiny portion of the total redshift, effectively by using the GW emission at farther distance.
	
    The second advantage for NSBH over BNS events comes from the absolute uncertainty in determining the mass ratio $q\equiv m_{s,2}/m_{s,1}$, which is smaller with $1/q\gg 1$ for NSBH events where the waveform is typically extended in time providing an accurate orbital definition. Consequently, this enables more accurate measurements of the 0.5PN signal on $q$ via monitoring the time evolution of the frequency spectrum (i.e.: chirping). As the uncertainty in $m_{s,2}$ is rather fixed by the intrinsic scatter of the NS mass function, the significant reduction in $\delta q$ implies a tighter constraint in the redshifted-chirp mass $(1+z)\mathcal{M}_{c}$, and consequently in $d_L$. 
	For BNS systems $q \approx 1$ and thus the mass ratio is relatively less well constrained by the data, leading to significant $H_0$ uncertainty, whereas the tight prior restriction from the NS mass function instead provides a desirable tight prior on the source frame chirp mass $\mathcal{M}_{c}$ for BNS events. 
	
    We will examine the difference between NSBH and BNS events in detail using simulation in Section~\ref{sect: event-level-uncertainty}.
	
	\section{Data} \label{sect: data}
	\subsection{NSBH Mass function in the Milky Way}

 The initial birth masses of neutron stars are predicted to lie in the range, $M_{\rm birth} \sim 1.08 - 1.57 M_\odot$, depending on the modeling of the underlying hydrodynamical processes and thermodynamics with subsequent evolution. 
  In the case of close binaries, as neutron stars accrete material from companions, leading to a dependence of the NS mass on the physical properties of the secondary, companion star. 
The upper limiting mass of a NS star is also understood to sensitively depend on uncertain nuclear reactions and also rotation rate affecting the equation of state \citep{ns-mass-function-compile}, under physically extreme nuclear conditions, allowing unique tests of the standard model in conditions as yet unrealized in the laboratory.

    \begin{figure}
	    \centering
	    \includegraphics[width=\columnwidth]{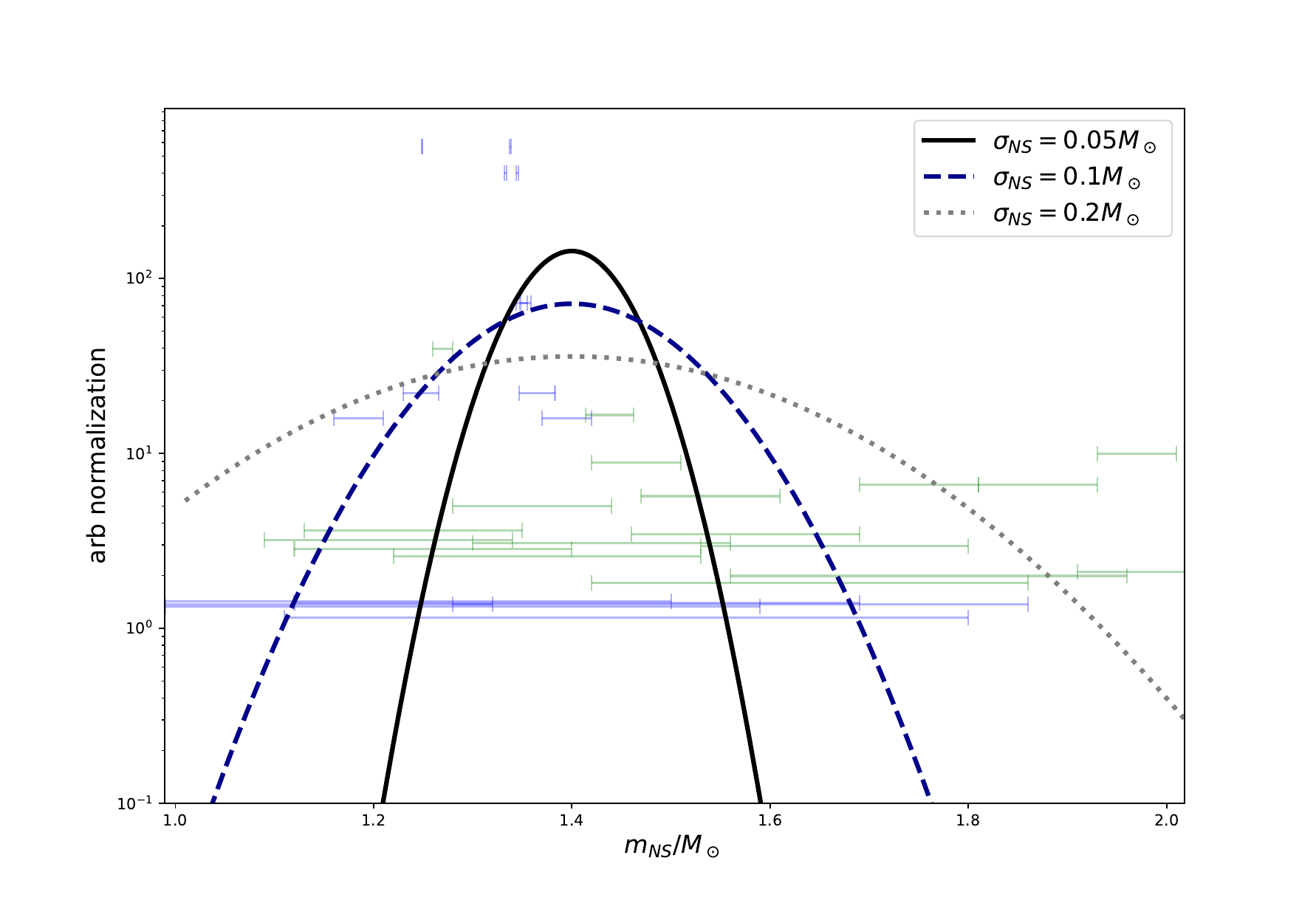}
	    \caption{The EM-observed masses of neutron stars in a double neutron star system (in blue) and neutron star-white dwarf systems (in green), compiled in \citep{ns-mass-function-compile}. We also plot the $3$ fiducial model of the underlying neutron star mass function - as a Guassian with a peak at $\langle m_{\rm NS} \rangle = 1.4 M_\odot$ and with various variance as shown in the legend.  }
	    \label{fig: ns-massfn-obs}
	\end{figure}
 
	Fortunately for our purposes the empirically determined NS mass function is well defined, with an inherently narrow spread of masses. Decades of observations have established that the majority of neutron stars have masses consistent with a single mass scale of $\langle M_{\rm NS}\rangle = 1.4M_\odot$, coinciding with the classically predicted for a universal iron core \citep{ns-eos-review}, including over $\sim 50$ radio pulsars  \citep{ns-mass-radio-pulsar1999}. Hence, we first adopt this empirically well defined prior mass function, approximately described simply by a Gaussian centered at $\langle M_{\rm NS} \rangle = 1.4 M_\odot$, and consider an intrinsic spread of NS mass ranging over $\sigma_{\rm NS} \in [0.05,0.2] M_\odot$ in our subsequent analysis below. A possible bimodal mass distribution	for Galactic neutron stars has been claimed in recent analyses for a classification by the type of binary companion,  \citep{ns-mass-function-compile} and fitted to two Gaussians with of differing mean mass and variance. For binary NS systems, a mean NS mass of $\langle M_{\rm NS} \rangle = 1.35 M_\odot$ is claimed to be significantly smaller the mean NS mass of NS-white dwarf binaries, estimated to be $\langle M_{\rm NS} \rangle = 1.5 M_\odot$ and with a smaller spread in mass. 
	Other classifications are also proposed \citep{ns-mass-function-companion-group2013} that do not separate companion types \citep{ns-mass-function-no-group} (see \citep{ns-mass-function-review} for a review of recent progress of constraining NS mass function), may also suggest two separated mass peaks with masses similar to \citep{ns-mass-function-compile, ns-mass-function-zhang-2011}. 
    We replot the EM-observed NS mass data from \citep{ns-mass-function-compile} in Figure~\ref{fig: ns-massfn-obs}, along with the $3$ fiducial models of the NS mass function that we would use throughout this analysis.
    We emphasis that our framework of analysis can be easily adopted to any NS mass function and in considering a binary mass distribution. 
    We find that the $H_0$ posterior inferred from such a simple Gaussian model does not significantly differ from adopting a more complicated NS mass function, so that our $H_0$ forecast is insensitive to the details of the NS mass function given the GW data quality anticipated. 
    This argument would be verified in Section~\ref{sect: forecast}.
	
	Lastly, there is also evidence from the gamma-ray that there can be some NS-like compact objects with relatively high masses \citep{ns-mass-upper-lim-xray}, with an upper extreme of $M_{\rm NS} \lesssim 2.3M_\odot$ established empirically recently for the
	well studied `black widow' pulsar PSR B1957+20 \citep{black-widow-1988-nature} (and a more recent analysis \citep{black-widow-review-romani-2022}). 
	The abundance of these relatively massive MS pulsars are much lower than the `classical' population at $\langle M_{\rm NS} \rangle \pm \sigma_{\rm NS}$ and these massive examples appear separated from the `classical' population tightly clustered around $1.4M_\odot$, allowing the possibility of excluding this uncertain, massive population from our analysis, so we may restrict ourselves to the predominant NS population both in calibrating the NS mass distribution and in comparison to the NSBH events observed from the GW channel. We anticipate that as long as we consistently exclude conspicuously massive NS related events that this prior selection will not induce a significant bias in our subsequent analysis. 
	
	\subsection{Gravitational Waves from NSBH events.}
	
	\begin{table*}
    \centering
    \begin{tabular}{|c||c|c|c|c|c|c|}
        \hline
        Event ID & $m_1(1+z) / M_\odot$ & $m_2(1+z)/ M_\odot$ & $\mathcal{M}(1+z)/ M_\odot$ & $q$ & $d_L/\mathrm{ Mpc}$ & Ref.\\
        \hline
        \hline
        \textbf{GW190426\_152155} & $6.45_{-1.78}^{+1.07}$ & $1.61_{-0.34}^{+0.24}$ & $2.60_{-0.006}^{+0.007}$ & $0.28_{-0.15}^{+0.05}$ & $394_{-109}^{+105}$ &   
        \begin{tabular}{c}
             \mycitet{gwtc2.1-catalog} \\
             \citet{mbta-search-gwtc2,gw190426-nsbh-or-not} 
        \end{tabular}
        \\
        \hline
        \textit{GW190917\_114630} & $11.2_{-2.46}^{+2.08}$& $2.45_{-0.50}^{+0.24}$ & $4.21_{-0.018}^{+0.014}$& $0.24_{-0.12}^{+0.03}$& $740_{-195}^{+173}$ & \mycitet{gwtc2.1-catalog}\\
        \hline
        \textit{GW191219\_163120}  & $33.2_{-3.61}^{+2.44}$ & $1.35_{-0.08}^{+0.07}$ & $4.81_{-0.045}^{+0.021}$ & $0.04_{-0.007}^{+0.005}$ & $682_{-168}^{+187}$ & \mycitet{lvk-gwtc3-data-paper}\\
        \hline
        \textit{GW200105\_162426} & $9.26_{-1.64}^{+1.07}$& $2.11^{+0.19}_{-0.28}$ & $3.62_{-0.007}^{+0.005}$ & $0.24_{-0.08}^{+0.04}$& $283_{-76.7}^{+79.6}$& \mycitet{lvk-gwtc3-data-paper,lvk-NSBH-discovery-paper} \footnote{$p_{\rm astro} = 0.36$}\\
        \hline
        \textbf{GW200115\_042309}  & $6.82_{-1.26}^{+1.13}$ & $1.48_{-0.25}^{+0.13}$ & $2.58_{-0.004}^{+0.004}$& $0.23_{-0.09}^{+0.04}$& $283_{-82.7}^{+62.1}$& \mycitet{lvk-gwtc3-data-paper,lvk-NSBH-discovery-paper}\\
        \hline 
        
    \end{tabular}
    \caption{List of candidate NSBH events. We list the reported observed frame quantities here, rather that the more model dependent source frame quantities as shown in the table compiled by LVK 
    \mycitep{lvk-gwtc3-data-paper}. The uncertainty ranges are the $68\%$ highest posterior density intervals. Remarkably, the uncertainty are correlated and clearly visible in terms of $q\equiv m_2/m_1$: where the error bar on this mass ratio is much smaller than $m_1(1+z)$ and $m_2(1+z)$. Values shown are only for quick reference; we use the full posterior samples in deriving these summaries, and similarly in all the subsequent analysis and only the events in \textbf{bold font} are regarded as reliable for the the analysis in this paper, in reference to the reported reliability of the classification of these events, as described in Section~\ref{sect: data}}
    \label{tab: gw-summary}
    \end{table*}
    The LIGO-Virgo-Kagra collaboration (LVK) completed a third observational run in March 2020 and released the GWTC-3 catalog \mycitep{lvk-gwtc3-data-paper}, containing $90$ observed binary coalescence events. The majority are classified as binary black holes mergers (BBH), with initially two events classified as neutron star-black hole mergers (NSBH) \mycitep{lvk-NSBH-discovery-paper}.
    Subsequently only one of these events GW200115 has survived scrutiny and now reported in the combined final catalogue\mycitep{lvk-gwtc3-data-paper}. An earlier event initially classed as BBH may now conform better to the NSBH definition, as we tentatively argue below, namely GW190426.
    
    The identification of NSBH events is predominantly based on the observed secondary mass $m_{o,2} = (1+z)m_{s,2}$, unlike BNS events. In principle, NSBH can produce a more complicated waveform than is BBH due to the tidal deformation of the NS, however the current data quality does not yet able to constrain such higher order tidal effects \citep{ns-love-hubble}. Therefore the claimed NSBH event(s) mostly rest on a low secondary mass estimated to 
    lie below a maximum of $m_{o,2} \lesssim 2.5 M_\odot$. 
    In Table~\ref{tab: gw-summary}, we list all the reported candidates events, including their posterior mean primary masses $m_{o,1} >5 M_\odot$ and secondary masses $m_{o,2} < 2.5 M_\odot$, comprising a total of $5$ events satisfying this criterion. These events are in general of relatively low SNR (as the SNR scales almost linearly with chirp mass), but is compensated to some extent by the better instrumental sensitivity at higher frequencies. GW events with $m_{s,2} = 2.6 M_\odot$ (GW190814) and $m_{s,2} = 2.8 M_\odot$ (GW200210) have been reported in GWTC-3. 
    Some of us have discussed these separately in the context of possible lensing of NSBH events that allows detection at higher redshift where the chirp mass is enhanced significantly by (1+z) \citep{lensed-nsbh-bds-2022}. 
    Furthermore, as will be shown in Section~\ref{sect: forecast}, the robustness of our method is not strongly influenced by the inclusion or the exclusion of such high mass events.
    
    In Table~\ref{tab: gw-summary} we indicate the status of the 5 possible NSBH candidate events reported to date. 
    (We would only quote the GW in the table by the first half name of the event, ignoring the ID behind the underscore.)
    Three of these are reported as marginal, including GW190426 with a high false alarm rate $32 \, \mathrm{yr}^{-1}$ in the MBTA pipeline, first reported in the GWTC-2.1-marginal catalog. (\citet{mbta-search-gwtc2}, \mycitet{gwtc2.1-catalog}) As commented by LVK \mycitep{gwtc2.1-catalog}, such a high false alarm rate is mainly attributed to the initially low NSBH event rate adopted by prior expectations based on the absence of NSBH events, that may be regarded as a somewhat circular argument resulting in the probability of an astrophysical origin assigned to this event of $p_{\rm astro} \leq 0.12$ across all the analysis pipelines. Hence, this `prior' may now be revised upward in retrospect with the subsequent confident identification of GW200115 as an NSBH event. The event GW191219 is also a marginal detection with a high false alarm rate of $4.0 \, \mathrm{yr}^{-1}$, and the secondary mass in the source frame is relatively low mass
    $m_{s,2} < 1.2 M_\odot$ at $68\%$ credibility. This marginal event is also with by far the most extreme mass ratio $q=0.04^{+0.005}_{-0.007}$ among the observed events, raising further doubt about its viability as real astronomical event and this is why it flagged as by the LVK as marginal (\mycitep{lvk-gwtc3-data-paper} Table II). 
    
    A third marginal NSBH candidate event is GW200105 of low SNR, partially due to LIGO Hanford was offline at the time of detection. The only detector that reached the desired SNR is the LIGO Livingston \mycitep{lvk-NSBH-discovery-paper}, where a chirping signal are clearly observed when the data stream are plotted on a Q-transform spectrogram.\mycitep{lvk-NSBH-discovery-paper}.
    Despite the relatively strong SNR$=13.9$ reported in \mycitep{lvk-gwtc3-data-paper}, the event has a $p_{\rm astro} = 0.36$. For this reason, this event is classified as marginal by the LVK in \mycitep{lvk-gwtc3-data-paper}. 
    
    
    This leaves two NSBH candidates, including GW200115, which has been consistently classed by the LVK as a secure NSBH event \mycitep{lvk-NSBH-discovery-paper}, whereas the earlier revised detection GW190426 which although not classed as NSBH or claimed as such does seem to satisfy the NSBH requirement in terms of the security of its detection and the reported component masses. 
    So we focus on these two events here and in particular, GW200115 given the secure status of this NSBH event as claimed by the LVT \mycitep{lvk-NSBH-discovery-paper}.
    
    \section{Statistical Framework} \label{sect: data-analysis}
    Throughout the paper we use the notation $\pi(\cdot)$ to denote the prior distribution, in contrast to the probability distribution $p(\cdot)$ to describe relations among the model parameters and the data.
    
    \subsection{$H_0$ posterior}
    We aim here to find the posterior for $H_0$ given the event waveforms, $\lbrace D_i \rbrace \equiv \bigcup_i D_i$, namely $p(H_0 | \lbrace D_i \rbrace)$. For each event with observed waveform $D_i$, the LVK collaboration has fitted a cosmology-free event-level posterior $p_i(d_L, z, \theta | D_i )$, with  $\theta$ are the remaining event level parameters (excluding $d_L$ and $z$). We wish to `sum' all the events properly to infer the (population-level) value of $H_0$. 
    
    To avoid \textit{ab initio} fitting, we use the following reweighing scheme as proposed originally for galaxy morphology dataset in \citep{Hogg10-pop}, and employed subsequently in GW analyses \citep{Farr19-pop}. 
    The posterior can be rewritten using Bayes theorem to flip $H_0$ as the condition, and expand the marginalized event level parameters as:
    \begin{align}\label{eq: h0-posterior-step1}
        p(H_0 | \lbrace D_i \rbrace) &\propto \pi(H_0)\int p(\lbrace D_i \rbrace | d_L, z, \theta) p(d_L, z, \theta |H_0) \, dd_L\,dz\,d\theta \\
        &= \pi(H_0) \prod_i^{N_{\rm obs}} \int p( D_i | d_L, z, \theta) p(d_L, z, \theta |H_0) \, dd_L\,dz\,d\theta .
    \end{align}
    The next step is to invoke the Bayes theorem to rewrite $p( D_i | d_L, z, \theta)$:
    \begin{equation} 
        p( D_i | d_L, z, \theta) = \frac{p(d_L, z, \theta | D_i, \rm{LVK} )p(D_i|\rm{LVK})}{p(d_L, z, \theta|\pi_{\rm LVK})},
    \end{equation}
    with $p(d_L, z, \theta|\pi_{\rm LVK})$ being an uninformative parameter prior adopted by LVK. 
    Following a similar notation, the term $p(d_L, z, \theta | D_i, \rm{LVK} )$ represent the parameter fitting results when the LVK uninformative prior is used to fit the data.  
    For a similar reason, the term $p(D_i|\rm{LVK})$ is an uninformative data prior, which is mathematically equal to a constant, and could be absorbed as an overall normalization constant.
    In general, these terms can have non-trivial form to account for the selection bias as well, and we will elaborate this in Section~\ref{sect: selection}. 
    
    Note the $K_i$ sample points of the posterior $p(d_L, z, \theta | D_i, \rm{LVK} )$ are available in the LVK data releases. Using these samples, one can rewrite Eq.~\ref{eq: h0-posterior-step1} as:
    \begin{equation} \label{eq: full-posterior}
        p(H_0 | \lbrace D_i \rbrace) \propto \pi(H_0) \prod^{N_{\rm obs}}_i \frac{1}{K_i}\left[ \sum^{K_i}_{\sim \theta, d_L, z}\frac{p(d_L, z, \theta | H_0)}{p(d_L, z, \theta|\pi_{\rm LVK})} p(D_i|\rm{LVK})\right] ,
    \end{equation}
    where the integration $\int d\theta dd_L dz$ is replaced by the Monte-Carlo sum $\frac{1}{K_i}\sum^{K_i}$.
    As argued above, every term except $p(d_L, z, \theta | H_0)$ in the square bracket can be modelled as an overall multiplicative constant.
    The data fitting results come into the above likelihood function via the Monte Carlo summation $\sum$, while the goodness of fit to the model is assessed in the term inside the Monte Carlo sum.
    In the subsequent subsection, we would explain how the goodness of fit term can be formulated.

    \subsection{Redshift-distance Likelihood}
     There is a non-trivial point about the MCMC samples released by LVK.
    Despite values of redshift $z$ are reported in the MCMC chain, those reported value of $z$ should be interpreted with extra care - indeed, their face values are not applicable to our application at all.
    
    The LVK fitting pipeline does not fit for the $(1+z)$ factor directly as it is completely degenerate with the restframe chirp mass Eq.~\ref{eq: redshifted-chirp-mass} in terms of the frequency response behaviour. 
    Instead, the LVK pipeline utilized the Planck cosmology \citep{planck2018-param} with $H_0 = 67.9$ km s$^{-1}$ Mpc$^{-1}$, so that the luminosity distance $d_L$ that parametrized the GW waveform is directly converted to the redshift, and these directly converted values of $z$ are reported in the MCMC chain.
    Therefore, the MCMC samples from LVK collaboration are samples of $\lbrace d_L, m_{o,2}, \theta \rbrace$ instead of directly containing $z$ 
    \footnote{There is a `no-cosmology' MCMC sample chain available in the LVK data release. However, the interpretation of `no-cosmology' is also different from what we are doing in this paper. In fact, their `no-cosmology' chain only means excluding a volumetric prior for event rate. The convention of assuming $h=0.679$ to convert $d_L$ to $z$ directly is still held in this `no-cosmology' chain. Therefore this chain cannot be used directly without using the likelihood marginalization method we presented in this paper. See Appendix \ref{app: lvk-likelihood-remark} for more detailed description for reproducing the results we shown here.}

    For this reason, all the probabilities containing $z$ in Eq.~\ref{eq: full-posterior} must be replaced by $m_{s,2}$, which indirectly constrain $z$ with the use of the Milky Way NS mass function. 
    
    Our prior here is the independent knowledge of the distribution of $m_{s,2}$ from the long standing Milky Way NS mass distribution, which we can factorize and write $H_0$ likelihood (the term inside the bracket in Eq.~\ref{eq: full-posterior}) as:
    \begin{equation} \label{eq: h0-model-likelihood}
    \begin{split}
        p(m_{o,2}, d_L,\theta|H_0) = \int dz \,dm_{s,2} \, d\beta \, &p(m_{o,2}|z,m_{s,2}) p(d_L|z,H_0) \\
        &\times p(z|\beta)  \pi(m_{s,2})\pi(\theta)\pi(\beta)
    \end{split}
    \end{equation}
    It is clear how the information regarding $z$ indirectly comes into the expression: if $m_{s,2}$ are known precisely, $z = m_{o,2}/m_{s,2}-1$ would inherit the data uncertainty in $m_{o,2}$. The above equation simply takes the intrinsic scatter of $m_{s,2}$ into account, promoting $m_{s,2}$ to a probability $\pi(m_{s,2})$. 
    
    Note the conditionals $p(m_{o,2}|z,m_{s,2})$ and $p(d_L|z,H_0)$ (i.e.: the first two terms before the multiplication symbol) are deterministic, and are thus possible to be expressed in terms of Dirac delta. This simplify the expression to:
    \begin{equation} 
    \begin{split}
        p(m_{o,2}, d_L,\theta|H_0) = \int d\beta \,  &p(z=z(d_L;H_0))|\beta) \\ 
        &\times \pi( m_{s,2} = m_{o,2}/(1+z(d_L;H_0)) ) \pi(\theta) \pi(\beta)
    \end{split}
    \end{equation}
    While the inverse function $z(d_L;H_0)$ does not admit an analytical form, a computationally inexpensive numerical solution can be easily obtained because $d_L(z;H_0)$ is a monotonic function of $z$. 
    When $d_L$ is known and a fixed value of $H_0$ is chosen in a MCMC sampling step, the corresponding value of $z(d_L, H_0)$ is uniquely determined. 
    And the likelihood of generating this specific $z(d_L, H_0)$ is given by the conditional $p(z=z(d_L;H_0))|\beta)$. 
    As the actual redshift prior $p(z|\beta)$ can be a complicated function that models both the unknown evolution of the intrinsic NSBH event rate and also the luminosity-limited selection effect, we will simply assume an uninformative prior. 
    Consequently the hyper-parameter $\beta$ to parametrize the redshift dependence is also trivial.
    In this way, the above model likelihood function would be using the NS mass function $\pi( m_{s,2} = m_{o,2}/(1+z(d_L;H_0)) )$ only.
    
        
    
    \subsection{Assessment of Selection Bias} \label{sect: selection}
    In a flux-limited survey, the observed luminosity distribution is always biased to the brighter end, so that even if the intrinsic NS mass distribution is symmetric, this selection effect skews the detectable mass distribution.
    However, as we will show below, this selection effect has only a negligible effect on our results. 
    
    Correction for selection bias in GW surveys is difficult, despite the existence of several attempts focused on the binary black hole systems \citep{bbh-selection-bias-2013,bbh-selection-bias-farr19,bbh-selection-bias}.
    In particular, the full treatment outlined in \citep{bbh-selection-bias-farr19} also incorporates the Poissonian count noise contributed by finite number of  observed events.
    In our analysis, we ignore the Poissonian noise in order to manage the computational cost.
    The selection effect for systems involving NS can be more subtle with the extra requirement of quantifying the tidal evolution of the NS during the NSBH merger.
    In the treatment quantifying selection bias below we also ignore all the extra contribution arisen from tidal effect. In other words, we treat neutron stars as essentially low mass black holes without any internal structure and deformability. 
    
    To correct for selection bias, the model term $p(d_L, m_{2,o}, \theta|H_0)$ in Eq.~\ref{eq: h0-posterior-step1} should be modified as the observed data segments $\lbrace D_i \rbrace$ must be passed through the selection criteria, therefore the model term should instead be understood as $p(d_L, m_{o,2}, \theta|H_0, 
    \lbrace D_i\in \textrm{det} \rbrace)$, and we have in fact implicitly included such condition in the likelihood $p(\lbrace D_i \rbrace | d_L, z, \theta, \lbrace D_i \in \textrm{det} \rbrace)$. 
    The importance of bias correction would thus be highlighted by the difference between $p(d_L, m_{o,2}, \theta|H_0,  
    \lbrace D_i\in \textrm{det} \rbrace)$ and $p(d_L, m_{o,2}, \theta|H_0)$.
    
    As $\mathcal{M}_{c}$ depends also on $m_{s,1}$, the correction for selection bias requires marginalizing over the uncertain black hole mass function $\pi(m_{s,1})$. 
    This can be factorized as:
    \begin{equation} \label{eq: selection-bias}
    \begin{split}
        &p(d_L, m_{o,2}, \theta|H_0,  
    \lbrace D_i\in \textrm{det} \rbrace) \\
    &= \int dm_{s,1} p(d_L, m_{s,1}, m_{o,2}, \theta|H_0,
    \lbrace D_i\in \textrm{det} \rbrace) \pi(m_{s,1})
    \end{split}
    \end{equation}
    The next step is to expand the term inside the integrand. Conceptually, we would like to cut the parameter subspace which is non-detectable, and then we normalize the remaining parameter space to ensure the unity of probability distribution. Symbolically:
    \begin{equation}
    \begin{split}
        &p(d_L, m_{s,1}, m_{o,2}, \theta|H_0,
    \lbrace D_i\in \textrm{det} \rbrace) \\
   &= \frac{1}{\mathcal{Z}} p(d_L, m_{s,1}, m_{o,2}, \theta|H_0 ) p(\lbrace D_i\in \textrm{det} \rbrace | d_L, m_{1,s} m_{2,o},\theta)
    \end{split}       
    \end{equation}
    
    \begin{figure}
	    \centering
	    \includegraphics[width=\columnwidth]{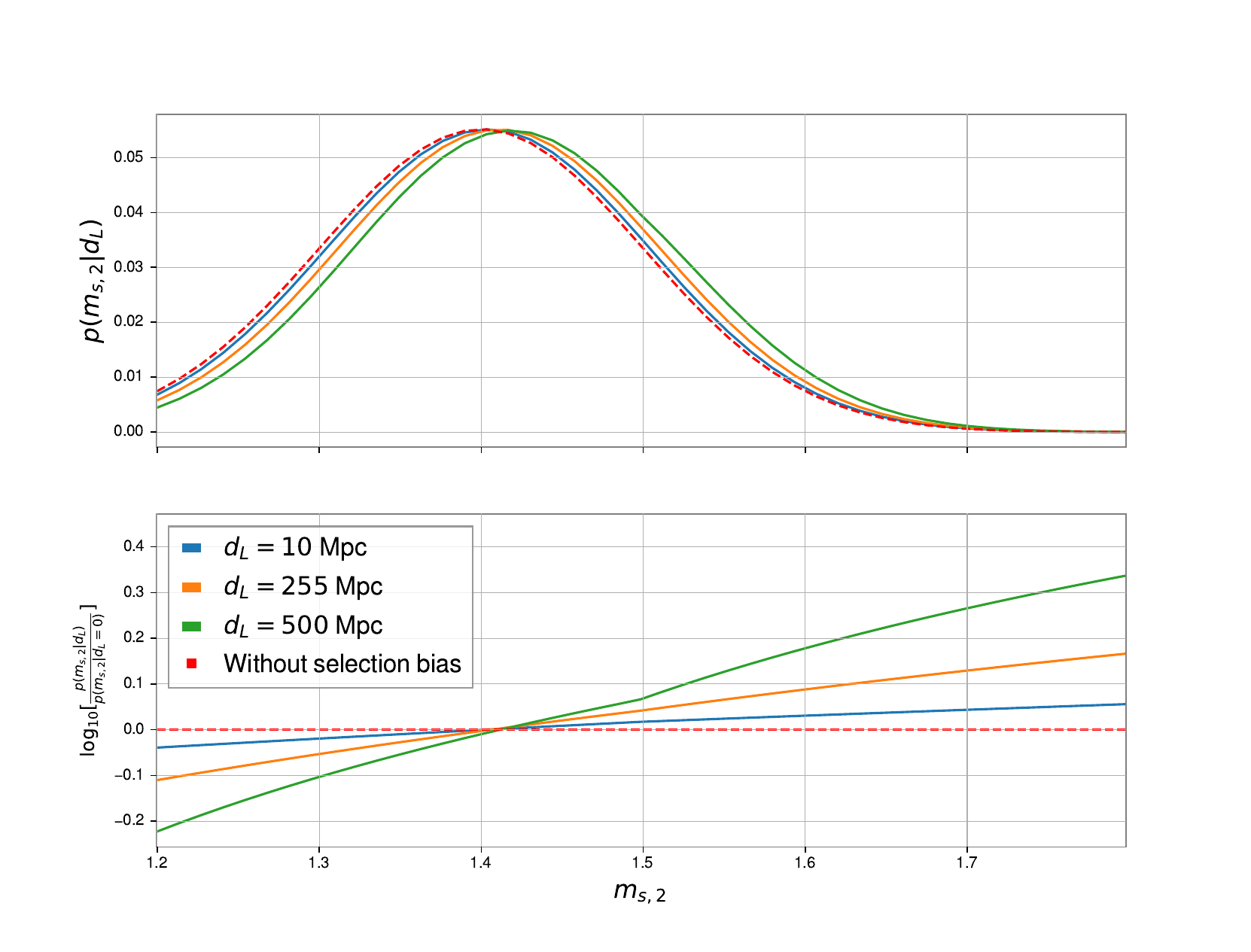}
	    \caption{Demonstration of the effect of selection bias. In the upper panel, we show the likelihood of finding mass $m_{s,2}$ at various observed distance $d_L$. In the lower panel, we show the fractional difference between the mass function with selection bias to the one without the bias (i.e.: the mass function evaluated at $d_L=0$).  } 
	    \label{fig: selection-bias}
	\end{figure}
    Here we model the detection probability $p(\lbrace D_i\in \textrm{det} \rbrace | d_L, m_{1,s} m_{2,o},\theta)$ using the code library \textsc{gwdet} \citep{gwdet}, in which the detection probability calculation are based on spline interpolation among the simulated injection samples defined on grids points of GW source parameters. 
    We model the biasing effect by $m_{s,1}$ by assuming the BH mass function $\pi(m_{s,1})$ follows the Gaussian + peak model as described in \mycitep{gwtc2-bbh-population,gwtc3-bbh-population}, although the detailed choice of $\pi(m_{s,1})$ in \mycitep{gwtc3-bbh-population} does not significantly change our result we will show.
    In Figure~\ref{fig: selection-bias}, we compare the expected NS mass function with and without corrected for selection bias. The effect of selection bias can be seen to increase with distance as a higher neutron star mass is necessary to produce sufficient SNR for event detection. 
    The red dotted curve shows the intrinsic NS mass function assumed in this comparison. 
    It is clear that for events in the nearby Universe that the observed NS mass function is not noticeably biased, and this is still the case even we consider events out to the detection limits of approximately $d_L \gtrsim 500$ Mpc, despite a slight potential skew towards the higher masses. Quantitatively, the peak NS mass is shifted only by $0.02 M_\odot$, which corresponds to $\sim 0.2-0.4 \sigma_{\rm NS}$. 
    We conclude that within the current detection horizon of NSBH event of approximately $d_L \approx 800$ Mpc, it is safe to ignore the modest level of selection bias, as the current data quality dominates the current uncertainty budget for $H_0$ estimated by our method.
     
    In this exploratory work we focus on demonstrating the feasibility of the proposed new measurement method for $H_0$ and for the rest of the paper, we drop the modelling of selection effect to keep the analysis simple, given the modest level of selection bias we have inferred as the analysis above suggests this selection effect does not have a significant role.
    Nevertheless we stress that the above scheme for modelling the selection bias can be employed in future analysis when larger samples of NSBH events are detected for future precision measurements of $H_0$ at the percent level.
    
    \subsection{Results: GW-190426 and GW-200115}
\begin{figure}
	    \centering
	    \includegraphics[width=\columnwidth]{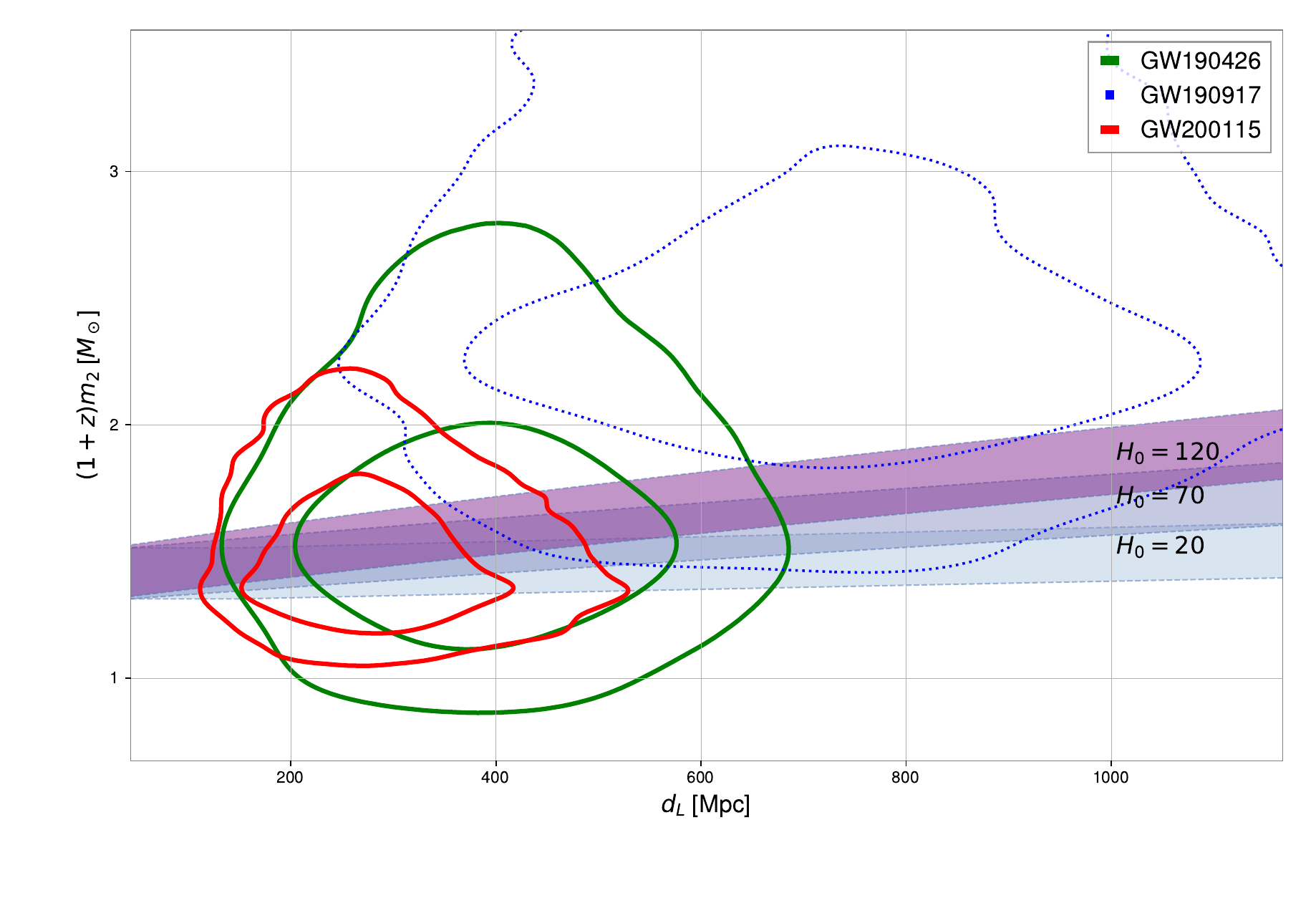}
	    \caption{Confidence contour the NSBH events listed in Tab.~\ref{tab: gw-summary}, overplotted with models that assume different values of $H_0$ (shaded bands). 
	   The y-position of the bands is determined by pinning down the the source frame NS mass $m_{s,2} = 1.4 \pm 0.1 M_\odot$, with the boundaries adopting $1.4+0.1 M_\odot$ and $1.4-0.1 M_\odot$ respectively  }
	    \label{fig: all-nsbh-event-posterior-H0}
	\end{figure}

    With the current limited number of observed NSBH events, inference of $H_0$ suffers from significant sampling effects, as the NS mass of these events may not fairly represent the underlying NS mass function.
    The purpose of this current analysis should be understood as a proof of concept, rather than a rigorous inference of $H_0$ from the existing GW data. 
    In particular, we focus on answering whether with the current signal-to-noise ratio, it is yet possible to deliver any meaningful constraint on $H_0$. 
    The main uncertainty currently, as will become clear below, is the distance estimate, $d_L$ derived from the event amplitude, rather than the observed mass $\mathcal{M}_c (1+z)$ for which the precision is relatively good for the current NSBH detections.
    In Figure~\ref{fig: all-nsbh-event-posterior-H0}, we show the $d_L$ verus $(1+z)m_2$ posterior for the NSBH events observed so far, with fiducial model assuming different value of $H_0$ overlaid on the posteriors.
    As can be seen, the $H_0$ constraint is substantially worsen by the intrinsic scatter of the NS mass function, and in general, the difference among different $H_0$ values start to become increasingly distinguishable at long distance $d_L$.
    This is also one reason why the use of NSBH is preferred over BNS system.
    In the plot we also show the posterior of GW190917 in dotted contour, which has an unusual high $m_{\rm s,2}(1+z)$ and is by now the most distant event with extreme mass ratio observed so far. 

\begin{figure}
	    \centering
	    \includegraphics[width=\columnwidth]{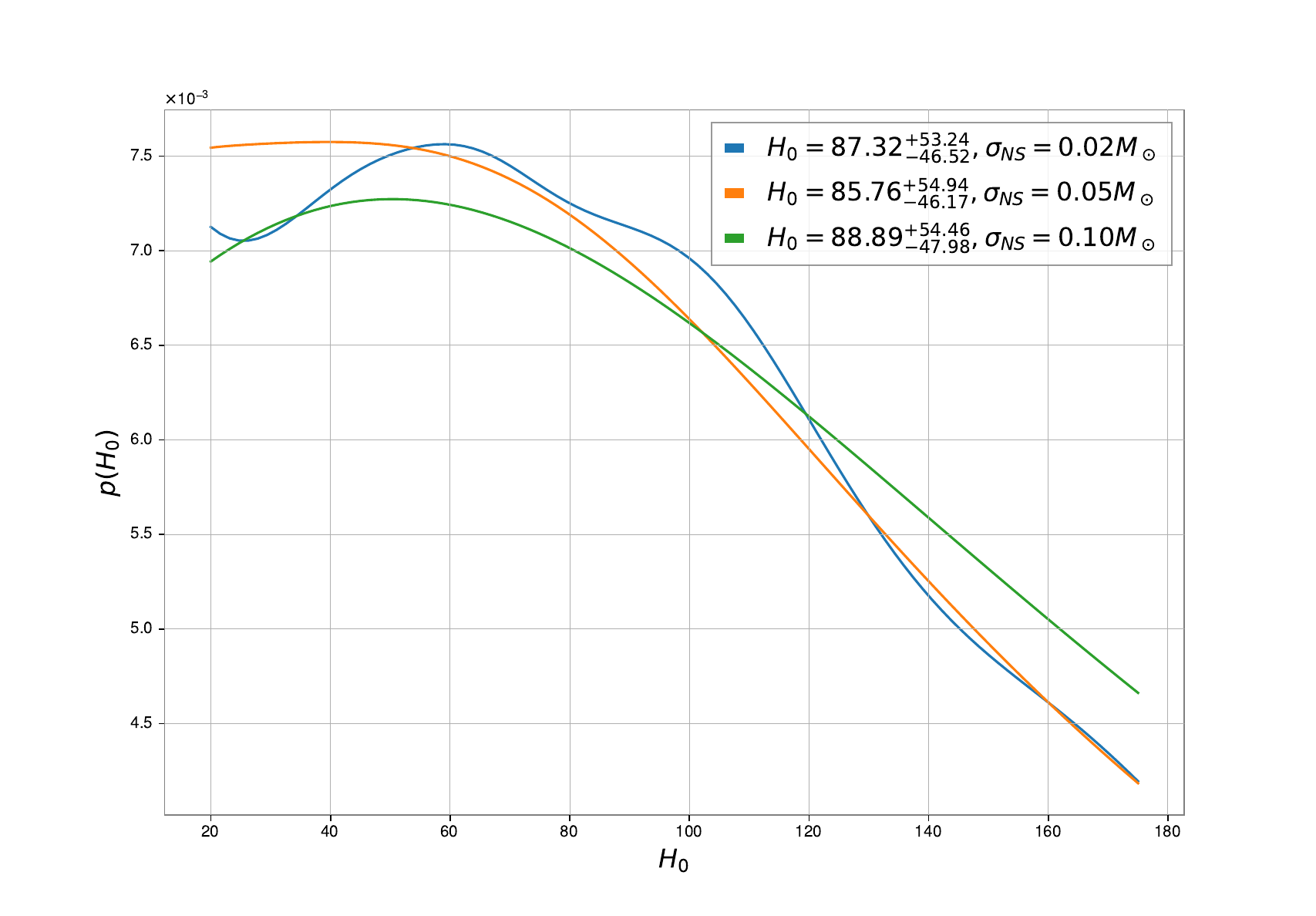}
	    \caption{$H_0$ constraint from the combined constraint of GW-190426 and GW-200115.}
	    \label{fig: commbined-event-H0}
	\end{figure}

    Using the more secure events as argued in Section ~\ref{sect: data}, namely GW190426\_152155 and GW200115\_042309, we perform the fitting procedure as outlined above to determine the recent constraint in $H_0$.
    As the scatter in NS mass function is still in debate, we repeat the analysis assuming $4$ characteristic value. 
    A fairly broad uniform prior of $H_0 \in [20, 180]$ km s$^{-1}$ Mpc$^{-1}$ is applied.
    The result is shown in Figure~\ref{fig: commbined-event-H0}.
    As the plot indicates, with the very few NSBH events and the poor event-level parameters recovery precision, the constraint on $H_0$ is not very strong, but a mild exclusion of $H_0 \gtrsim 130$ km s$^{-1}$ Mpc$^{-1}$ can be noticed.
    This is inline with the intuition from Figure~\ref{fig: all-nsbh-event-posterior-H0}. where the event posterior contours spread different models assuming different $H_0$ values. 
    Nonetheless, this demonstrate the applicability of our method.
    It is promising to notice the method can still deliver some exclusion to $H_0$ models even with the scarcity of data, which are expected to gradually improve in the near future. 
    A forecast of the improvement in precision, and a study of the systematics will follow in Section~\ref{sect: forecast}.

    \section{Forecasting Near Future Precision on $H_0$} \label{sect: forecast}
    As the observed number of NSBH events is still limited we now present a forecast to predict the achievable precision on $H_0$ that can be anticipated in the near future. This is estimated by a dedicated waveform injection pipeline. 
    We perform simulations of NSBH events drawn from an assumed event distribution, and examine how the waveform fitting procedure induces uncertainty in addition to the intrinsic scatter on the NS mass distribution adopted in our prior.

    The required modelling for the population of simulated events comprises three distributions, namely: the BH mass function $\pi(m_{s,1})$, the NS mass function $\pi(m_{s,2})$ and the event rate $\pi(z)$. We randomly draw samples from these distribution, and subsequently shifted the source frame quantity $m_{s,1}$, $m_{s,2}$ to the observed frame quantity $m_{o,1}$ and $m_{o,2}$. 
    The $\lbrace m_{o,1}, m_{o,2}, z \rbrace$ samples are then filtered to rule out events that generate weak signal undetectable before any waveform evaluation to cut down the computational cost. 
    Afterwards, we calculate the waveform parametrized only by \textbf{observed frame quantities} $\lbrace m_{o,1}, m_{o,2}, d_L \rbrace$ using the publicly available package \textsc{bibly} \citep{bibly-software,bibly-validation,bibly-sampler}, which is the default analysis package for LVK \mycitep{lvk-NSBH-discovery-paper} to analysis NSBH event. In particular, we use the approximant \textit{IMRPhenomXPHM} to better handle the extreme mass ratio for NSBH event. 
    We then project the waveform to the detectors LIGO L1, H1 and Virgo, and inject noise realizations to the projected waveform from their corresponding power spectral density. 
    These procedures create simulated observations, namely time series of strain variations among each of the detectors.
    
    These time series are then passed to our fitting pipeline. Our goal is to obtain the $H_0$ posterior using the simulated time series via Eq.~\ref{eq: h0-posterior-step1}. 
    However, as will be explained in the next subsection, we would leave flexibility in choosing the presumed population parameters characterising $\pi(m_{s,2})$. Therefore, the time series would first be fitted assuming an uniform prior in $m_{s,2}$ and $z$ via the MCMC sampler \textsc{dynesty} \citep{dynesty,bibly-sampler}. 
    We have the freedom to enforce different mass priors $\pi(m_{s,2})$ when the MCMC samples are subsequently used to calculate $H_0$ posterior, by re-weighing the MCMC samples as explained in Eq.~\ref{eq: full-posterior}. This approach effectively reduces the undesirable repetitions in MCMC sampling when we switch between priors, and thus cut the computational cost down to a affordable amount. 
    In the fitting procedure, we use the same likelihood function \citep{gw-param-accelerate} as adopted in LVK analysis \mycitep{lvk-gwtc3-data-paper}. It is numerically implemented in \textsc{bibly}, with the phase parameter being analytically marginalized over as demonstrated in \citep{gw-param-accelerate}. 
    
    \subsection{Event level uncertainty} \label{sect: event-level-uncertainty}
    While the event level parameter recovery and the corresponding uncertainty estimates from noisy signal as been extensively demonstrated by LVK \mycitep{lvk-gwtc3-data-paper}, most of the existing data are on binary BH signals. 
    For our application on NSBH, we would like to focus more on the parameter degeneracy related to $z$ and $d_L$, and how the correlations varies as switching between different injected event parameters, thus elucidate the limit on $H_0$ precision due to parameter degeneracy.
    
    \subsubsection{Parameter degeneracy}
    In Section~\ref{sect: framework}, our estimation shows the constraint on $z$ and $d_L$ are inferred from the two (almost independent) observables, namely the chirping spectrum (constrain $q$ and thus $(1+z)m_2$) and the signal amplitude respectively.
    It is therefore natural to investigate the shape of the posterior given the waveform data: whether the $z$ and $d_L$ are minimally correlated.
    In Figure~\ref{fig: bns-vs-nsbh}, we plotted the confidence contours on the $d_L$ and $(1+z)m_2$ plane. 
    If the two parameters are inferred from independent observables, the semi-major and semi-minor axis of the contours should be parallel to the $d_L$-axis and $(1+z)m_2$-axis respectively. \footnote{If the quantites are inferred from correlated observables, the quantites would consequently be linear combinations of all observables. Thus, the covariance matrix would possess off-diagonal terms that tilt the contours at some angle to the $d_L$-axis and $(1+z)m_2$ axis. }
    Indeed, our intuition is valid. Because of the absence of significant correlation between $(1+z)m_2$ and $d_L$, we can determine $z$ once $m_2$ is constrained in the NS mass function defined in the prior. 
    The independent measurement of $z$ and $d_L$ allows one to determine $H_0$ via the classic Hubble diagram. 
    
    \subsubsection{BNS vs NSBH} 
    \begin{figure}
	    \centering
	    \includegraphics[width=\columnwidth]{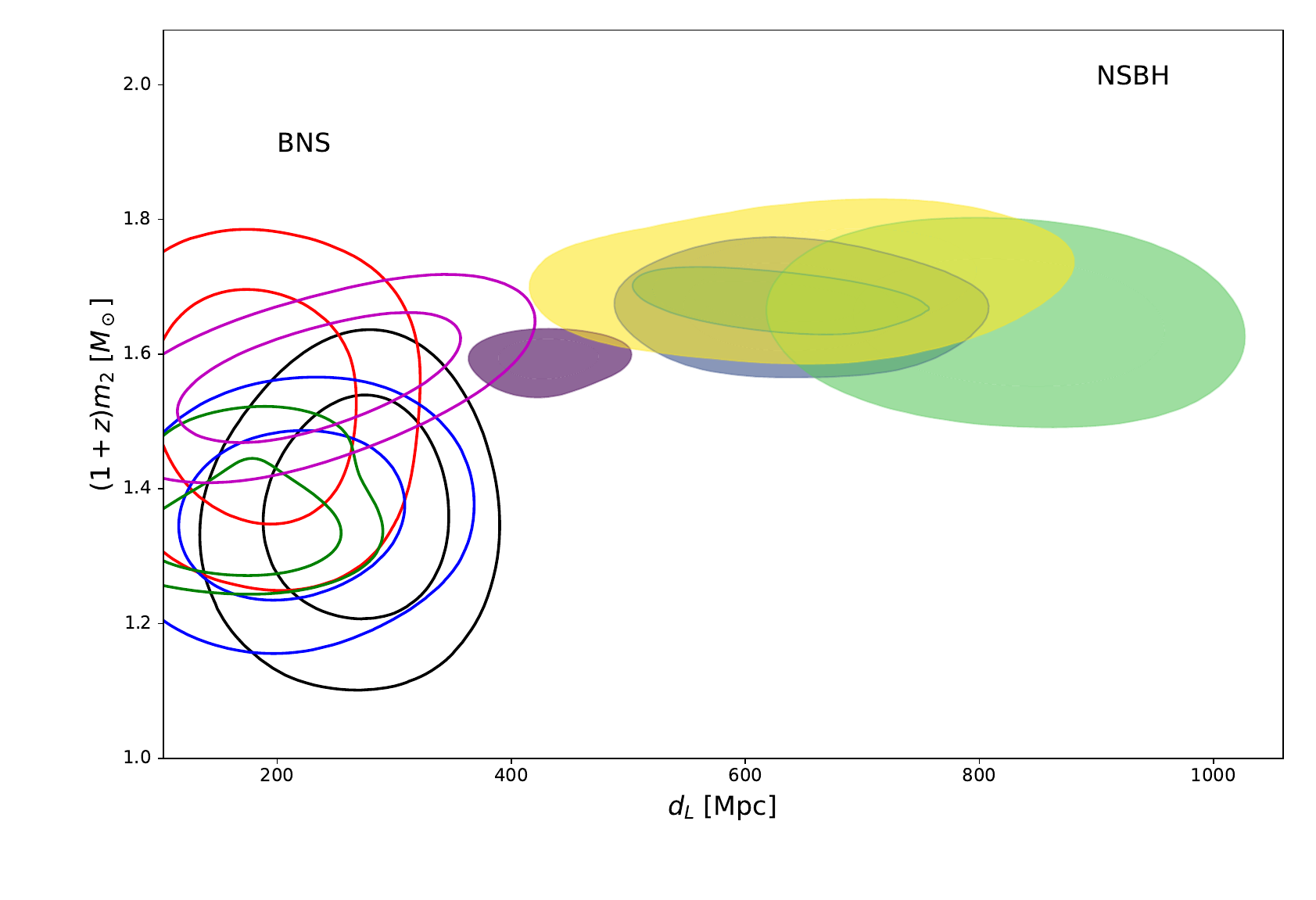}
	    \caption{The posterior constraint on $5$ simulated BNS (unfilled contours) and NSBH events (filled contours) respectively. Clearly, despite the uncertainty in $d_L$ is slight larger for NSBH, the uncertainty in $(1+z)m_2$ is substantially reduced when compared to BNS event. Therefore a better precision on event redshift $z$ are archived by using NSBH event. }
	    \label{fig: bns-vs-nsbh}
	\end{figure}
	In particular, we would like to understand the advantage of using NSBH over the BNS event to a deeper detail. 
	As argued previously in Section~\ref{sect: framework} and Eq.~\ref{eq: h0-uncertainty-fisher}, due to the efficient amplification powered by the BH companion, NSBH provides the $\approx 4$ enlargement of the detection horizon, and thus constrains $H_0$ much better than BNS. 
	
	Recall the precision on constraining $H_0$ also depends on the estimation uncertainty in $z$. 
	Apart from the extension of the detection horizon, NSBH also provide tighter estimate on $z$. Mergers with extreme mass ratio generates waveform that is elongated in time, and thus provide more sample points to improve the estimation on the chirping spectrum which constrains $q$. 
	To test this intuition, we inject a population of BNS events with both the component masses sampled from the same NS mass function $\pi(m_{s,2})$, and assign the heavier component as $m_1$. Similar to what we did for the NSBH events, we fit the injected waveforms using a flat mass prior for both the masses, but adopting the approximant for BNS \textit{IMRPhenomPv2-NRTidal}. The information of the NS mass function is used only when inferring the $H_0$ posterior (Eq.~\ref{eq: h0-model-likelihood}). 
	Such choice would allow fair comparison among the two types of sources.
	
	The recovered posterior on $d_L$ and $(1+z)m_2$ for each events are plotted in Figure~\ref{fig: bns-vs-nsbh}.
	As can be seen, the detectable BNS mostly fall in the more nearby universe, and the associated estimation of $(1+z)m_2$ are more uncertain. 
	In contrary, most NSBH injected are originated at a larger distance as the sampling volume increases. 
	However, the estimation uncertainty in $d_L$ does not worsen significantly. Furthermore, there is a notable improvement in constraining $(1+z)m_2$ and thus $z$. 
	This demonstrates the advantages for using NSBH over BNS.

    \subsection{Population level uncertainty}
    Another limit on precision determination of $H_0$ comes from (1) the intrinsic scatter of the source population, and (2) the finite sampling from the source distribution. 
    
    \subsubsection{Sensitivity to source distribution}
    To examine effect (1), we generate events from $\pi(m_{s,2}; \sigma_{\rm NS})$ assuming different level of intrinsic scatter $\sigma_{\rm NS}$, and feed the simulated data to our fitting pipeline to obtain the posterior $p(H_0 | \lbrace D \rbrace)$ following Eq.~\ref{eq: full-posterior}. Note we do not enforce to take the same intrinsic scatter $\sigma_{\rm NS}$ in the $H_0$ fitting procedure as in the simulation procedure, hence allowing tests of the sensitivity against incorrect characterization of $\sigma_{\rm NS}$. 

\begin{figure}
	    \centering
	    \includegraphics[width=\columnwidth]{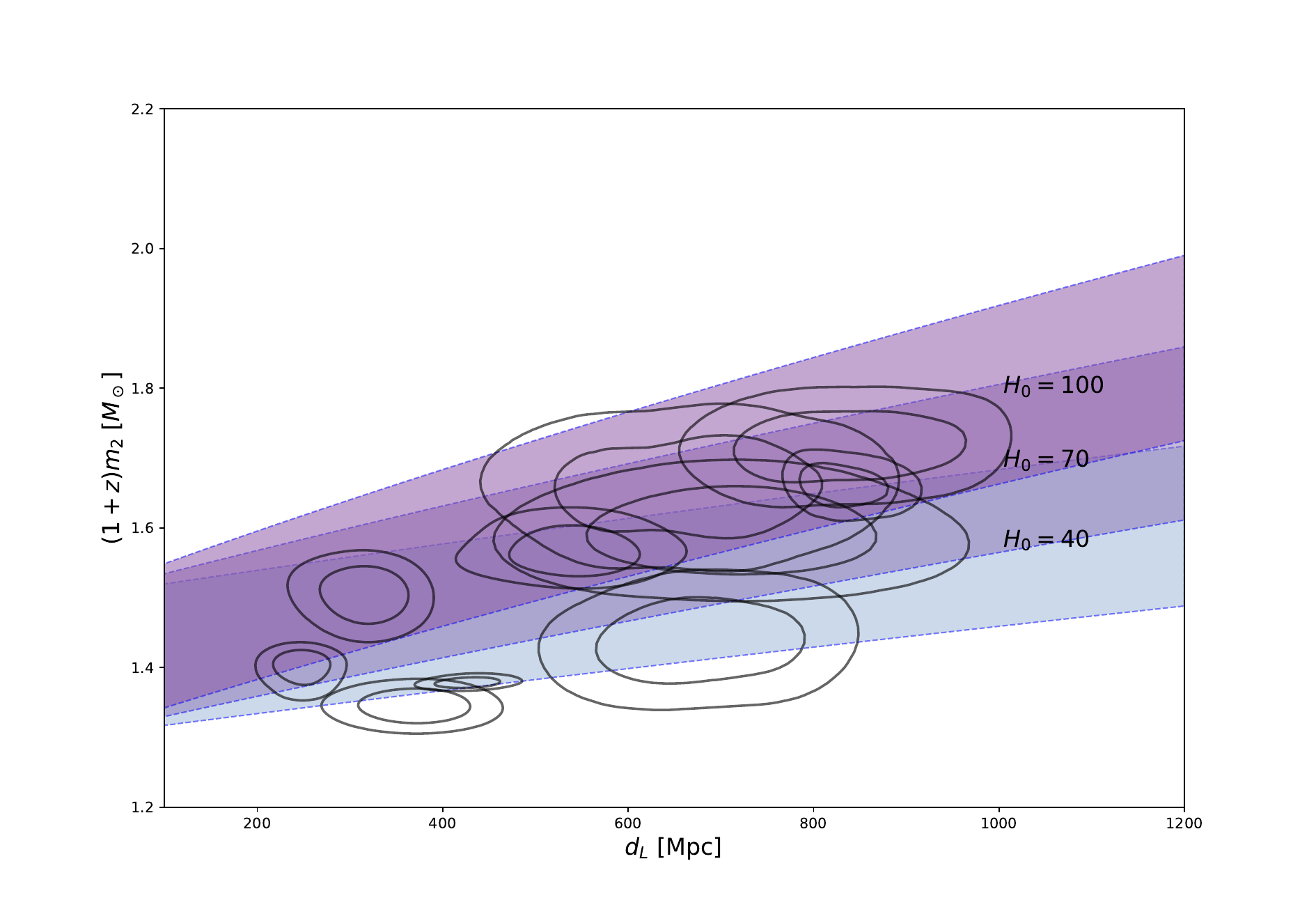}
	    \caption{The simulated Hubble Diagrams for the $10$ injected events, assuming $\sigma_{\rm NS} = 0.1 M_\odot$.}
	    \label{fig: simulated-hubble-diagram-10events}
	\end{figure}

    We injected a population of NSBH with $\langle m_{\rm NS} \rangle = 1.4 M_\odot$ and $\sigma^{\rm (sim)}_{\rm NS} = 0.1 M_\odot$; the large intrinsic scatter of $\sigma^{\rm (sim)}_{\rm NS} = 0.1 M_\odot$ is probably a conservative estimation. 
    We sampled $10$ events from the simulation, and determine the $H_0$ posterior assuming $\sigma_{\rm NS}^{(\rm fit)} \in \lbrace 0.02, 0.05, 0.1, 0.2 \rbrace M_\odot$.
    The Hubble diagram of these simulated events are shown in Figure.~\ref{fig: simulated-hubble-diagram-10events}.
    
\begin{figure}
	    \centering
	    \includegraphics[width=\columnwidth]{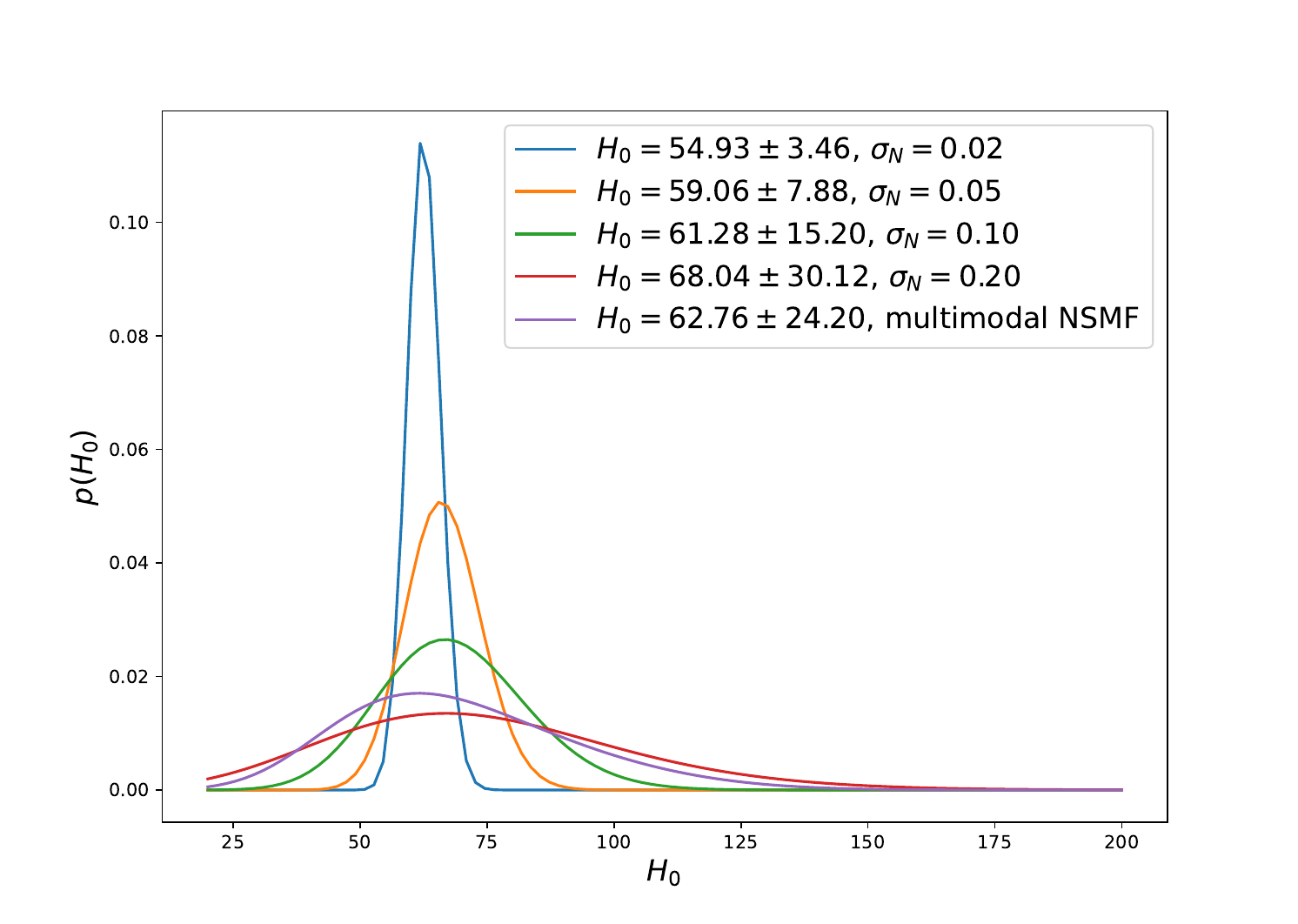}
	    \caption{The constraint in $H_0$ when $10$ simulated NSBH events are observed, assuming $\sigma_{N}^{(\rm sim)} = 0.10 M_\odot$. The multimodal neutron star mass function (marked as NSMF on the plot) assumes the two peaks at $m_2 = 1.3$ and $1.5$ $M_\odot$ respectively, with a equal dispersion on each peak $\sigma=0.1 M_\odot$ and equal relative abundance. }
	    \label{fig: simulated-unceratinty-10events}
	\end{figure}

    It is clear from Figure~\ref{fig: simulated-unceratinty-10events} that an under-estimated $\sigma_{\rm N}^{\rm (fit)} < \sigma_{\rm N}^{\rm (sim)}$ would not only lead to the incorrect shrunk of error, but can in principle, substantially shift the mean $\langle H_0 \rangle$ to incorrect value, depending on the sampled mean NS mass $\langle m_{\rm NS} \rangle^{(fit)}$ in contrast to the simulated mean NS mass $\langle m_{\rm NS} \rangle^{(sim)}$. An over-estimate of $\sigma^{(\rm fit)}_{\rm NS} > \sigma^{(\rm sim)}_{\rm NS}$ tends to diminish such bias effect in the center value of $H_0$, but would exaggerate the uncertainty in $H_0$. 
    
	\subsubsection{Source distribution: multimodal}
	
	It has been reported that the NS mass distribution is more complicated than a simple Gaussian: instead of a peak around the Chandrasekhar limit $m_{2,s} \sim 1.4 M_\odot$, the NS mass function have (at least) two neighboring peaks at $m_{2,s} \approx 1.37 M_\odot $ and $1.57 M_\odot$ \citep{ns-mass-function-zhang-2011} despite the exact location differs in different literatures, depending on the selection of NS samples (see a summary in Section~\ref{sect: data}, or from \citep{ns-mass-function-review} and reference therein).
	
	
	As can been seen in Figure~\ref{fig: simulated-unceratinty-10events} (the purple curve therein), the presence of multiple peaks in the NS mass function would in general, skew the distribution towards the direction that corresponds to the less dominant peak. 
	Having said that, the mean value of $H_0$ can still be recovered. 
	If the peaks in the intrinsic NS mass function are not well separated due to intrinsic scatter, the corresponding $H_0$ posterior would also hide the peaks due the propagated uncertainty in $H_0$.
	

    \begin{figure*}
	    \centering
	    \includegraphics[width=\textwidth]{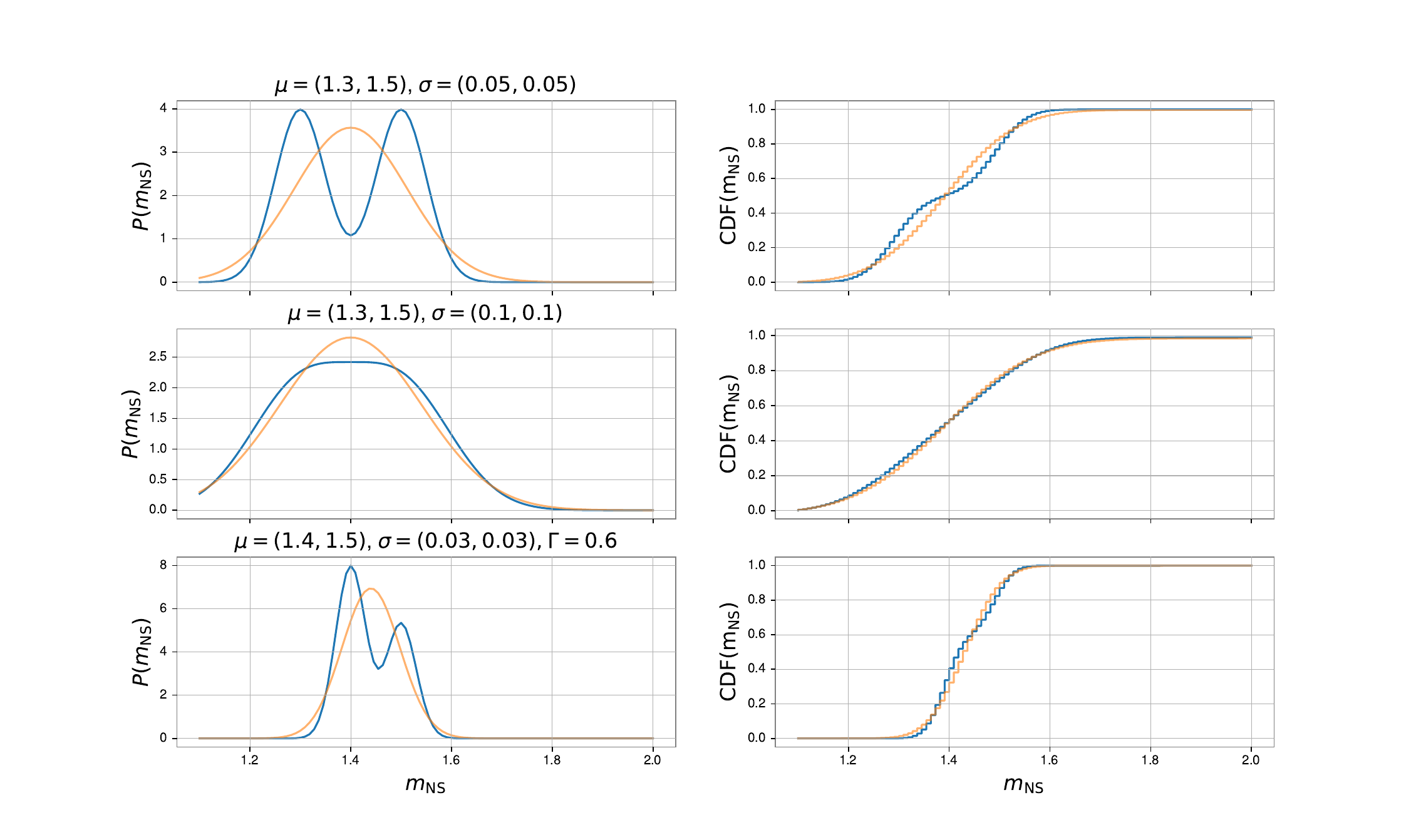}
	    \caption{We show the significance of the possible multi-modal feature as reported from some EM observation, and the approximation error for describing the multi-modal distribution by a single effective Gaussian. Three sets of multi-modal parameters are shown in blue line, with the corresponding effective Gaussian plotted in orange. The multi-modal distribution parameters are described in the plot title. On the right column, we show the same comparison, but in terms of the cumulative distribution function (CDF), which are traditionally use to define the distinguishability between distributions (via a Kolmogov-Smirnov test). As can be seen, in terms of the CDF, the effective Gaussian description is highly indifferent from the full, multi-modal description as long as the multi-modal peaks are not largely separated. }
	    \label{fig: multi-peak-compare}
	\end{figure*}
	
	The reasons for the striking similarity between the multimodal mass function and the simple Gaussian distribution can be understood as follows. The mean of a two-Gaussians mixture $\Gamma \mathcal{G}(\mu_1,\sigma_1)+ (1-\Gamma) \mathcal{G}(\mu_2, \sigma_2)$, with a relative abundance of $\Gamma \in [0,1]$, is $\Bar{\mu} = \Gamma \mu_1 + (1-\Gamma)\mu_2$. Similarly, the associated variance is $\Bar{\sigma}^2 = \Gamma (\sigma_1^2 + \mu_1^2) + (1-\Gamma) (\sigma_2^2 + \mu_2^2) - \Bar{\mu}^2$. 
	We can statistically assess whether it is possible to significantly distinguish a Gaussian mixture with an effective Gaussian $\mathcal{G}(\Bar{\mu}, \Bar{\sigma})$ via a two-sample Kolmogorov-Smirnov (KS) test.
	In particular, we focus on the limit that the number of samples from those distribution are finite, so as to mimic the fact that we would observe limited number of NSBH events. 
    As an visual comparison, we plot a few multimodal mass function with benchmark parameters chosen to mimic some of the claimed form in \citep{ns-mass-function-review}. 
    The results are shown in Figure~\ref{fig: multi-peak-compare}. 
    While the appearance of the effective Gaussian description may look very different in terms of the probability distribution function, they look fairly similar in terms of the cumulative distribution function (CDF) - based on which the KS test is built. 
    In particular, the difference is the most noticeable when $\Delta \equiv (\mu_2 - \mu_1) /\mathrm{min}(\sigma_1,\sigma_2) \gg 1$, so that the $2$ peaks can be `resolved'. 
    
    For realistic constraints from the MW observation of NS binaries, $\Delta < 1$, and thus the two description are statistically indistinguishable. 
    This conservative assessment did not take into account of the substantial GW measurement uncertainties in the redshifted NS masses $m_{\rm s,2}$; One would therefore expect the measurement uncertainties would further diminish the difference between the multi-modal distribution and the effective Gaussian.
    This result is indeed expected: in the effective Gaussian formalism, we force the corresponding mean and variance (the 2 leading statistical moments) to be consistent with the multimodal distribution. 
    Thus, any distinguishable feature would be contributed only starting from the 3-rd order moment, where the amplitude is suppressed at least by $\left[ (x-\Bar{\mu})/\Bar{\sigma} \right]^3 \ll 1^3$. 
	
	This demonstrates our method is insensitive to the fine detail of the underlying NS mass function, and is therefore robust to the selection of the samples from the electromagnetically observed neutron stars.
	
    \subsubsection{Dependence on number of events}
    Intuitively when the events are completely independent, the uncertainty in $H_0$ is reduced with increased numbers of events: $\delta^{(N)} H_0 = \frac{1}{\sqrt{N}} \delta^{(1)} H_0$. 
    However, we emphasize that the events are not independent despite the fitting procedures do not rely on the knowledge of other events. 
    The data generation process links the $m_{s,2}$ of each event together via $\pi(m_{s,2})$: the presence of a subset with $m_{s,2} < \langle m_{\rm NS} \rangle$ implies the likely presence of another subset with $m_{s,2} > \langle m_{\rm NS} \rangle$, so that the population mean $\langle m_{\rm NS} \rangle$ is probabilistically restored. The presence of this correlation makes the simple $1/\sqrt{N}$ scaling too optimistic. 
    
\begin{figure}
	    \centering
	    \includegraphics[width=\columnwidth]{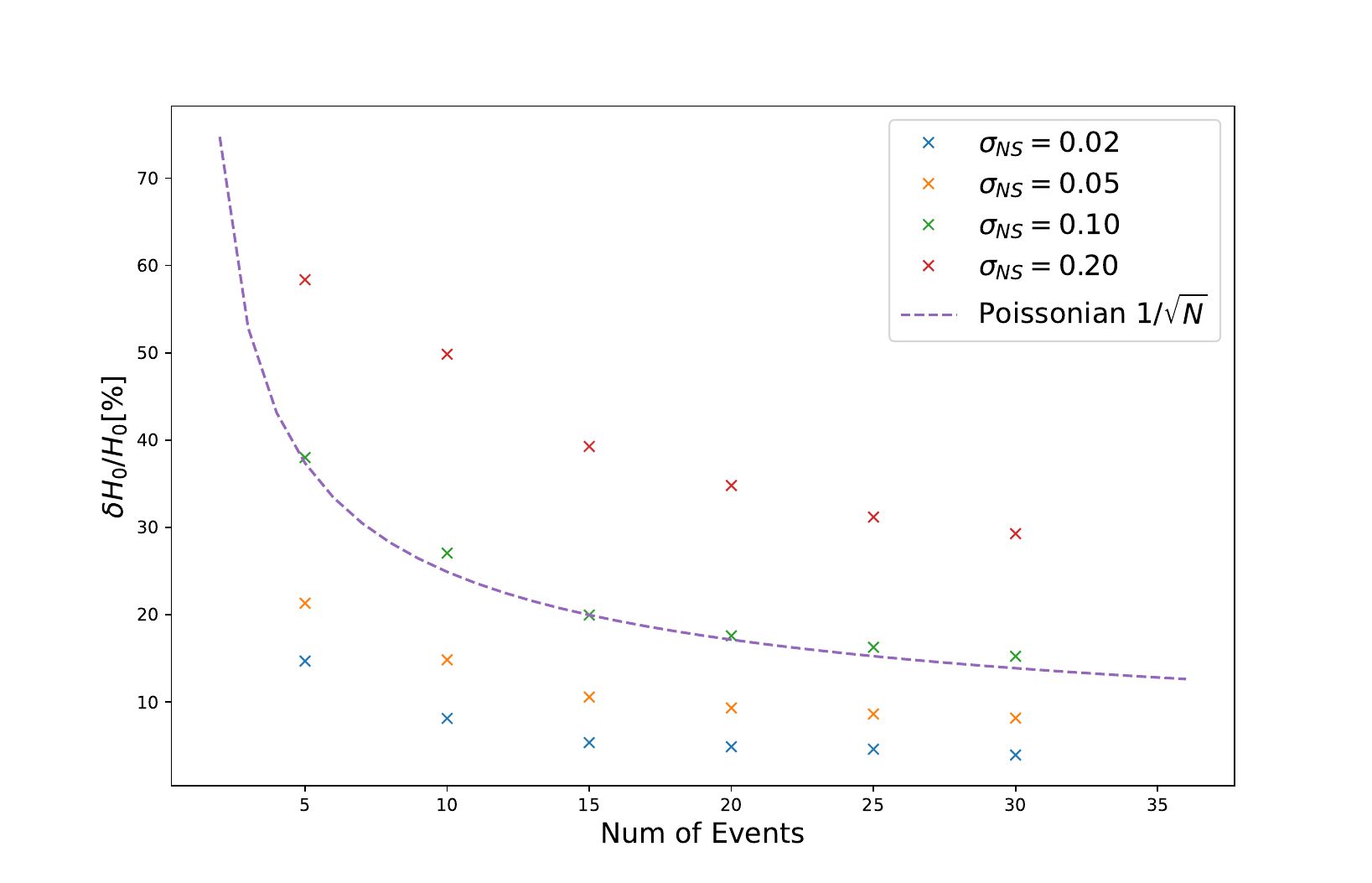}
	    \caption{The shrinkage forecast in percentage error in $H_0$ as a function of number of observed events. It is clear that the shrink in uncertainty improves slightly slower than the Poissonian decay, as the events are not totally indepedent: they share the calibration parameters (i.e. intrinsic properties of the mass function) in our fitting process.}
	    \label{fig: simulated-uncertainty-improvement}
	\end{figure}

    To study the impact of those possible hidden correlations, we re-use the simulated events and evaluate the posterior Eq.~\ref{eq: full-posterior} by supplying randomly sampled subsets of simulated events, with each subset containing different number of events. The results are shown in Figure~\ref{fig: simulated-uncertainty-improvement}. 
    It is clear that the improvement is slightly slower than $1/\sqrt{N}$, yet the improvement is still significant, so that for example the accumulation of $10$ events successfully shrinks the error to $1/2$ of the that from a single event, meanwhile driving the posterior mean $\langle H_0 \rangle$ to match the injected value. 
    
    \subsection{Discussion} 
    
    As we shown above, shrinking down the $H_0$ uncertainty innate to this method requires more detected NSBH events, which would be achievable in the near future.
    Apart from waiting for more observed NSBH-GW events, other efforts on EM-based observation of the Milky Way NS systems could also help improving the $H_0$ constraint. 
    One dominant uncertainty is inherited from the estimation of $z$ from the GW data, which require the external, `calibration' input from the neutron stars observed in the Milky Way.
    As explained in Section~\ref{sect: framework} (and the associated schematic Figure~\ref{fig: framework-summary}), once the neutron star mass $m_{\rm s,2}$ is determined, it could be combined with the spectral evolution of the GW events to determine the mass ratio $q$ and thus the source frame chirp mass $\mathcal{M}_{\rm c}$. 
    Such source frame chirp mass $\mathcal{M}_{\rm c}$ can then be compared with the redshifted chirp mass $\mathcal{M}_{\rm c}(1+z)$ to obtain an estimate of $z$. 
    On the other hand, there is an independent method to construct a data-driven, exteraal prior on the mass ratio $q$ as well.
    With an external estimate of $q$, the uncertainty in the GW spectral evolution constraint on $q$ would significantly shrink. 

    The EM measurement of mass ratio of binary system involving neutron star is not generally possible, unless at least of the one star of the binary system is a pulsar (or other time-transients).
    This could be understood as follows: astrometry measurements are able to measure the Keplerian parameters, which constraints the component masses up to an unknown orbit inclination.
    Such parameter degeneracy can be removed by a companion pulsar - exploiting the relativistic time delay when pulsar signal is deflected by the gravitational field. (See for example, \citet{ns-eos-review} for a review.)
    Indeed, in the neutron star mass function we picked as demonstration \citep{ns-mass-function-review, ns-mass-function-no-group, ns-mass-function-compile}, the neutron star mass are obtained from system that hosts at least a pulsar. 
    The measurements of $q$ for each systems allow the construction of the distribution of $q$. 
    Such externally constraint distribution can then be input as a prior to our analysis pipeline.

    Another subtle aspect about the prior on $q$ is to handle its potential correlation with the mass scale $m_{\rm NS}$ of the corresponding binary system.
    This is theoretically expected as the formation mechanism of NS with viable mass can be different - possibly including tidal interaction within the binary system where $q$ would be an important control parameter.
    Owing to limited amount of available of EM-detected NS system, such effect is still unclear; it is also for this reason we stay conservative in this analysis and do not model such effect.
    However, this should be of practical interest in the near future with the consortium of EM observation and GW observation to lift up such degeneracy, thus help tighten the constraint of $H_0$.

    A final remark on the application of BNS system, verus the NSBH system as discussed in the current work.
    As the neutron star mass function has been fixed in \textit{a priori}, in principle the expected chirp mass $\mathcal{M}_c$ can also been fixed by convolving the neutron star mass function with itself.
    Such process naturally lead to a prior directly on the chirp mass $\mathcal{M}_c$ for BNS system. 
    This is different from the NSBH approach we discussed here - where the primary mass $m_{\rm s,1}$ (black hole mass) and equivalently the mass ratio $q$ has to be constrained by the data.
    It should be now clear that the BNS approach would strongly influenced by the presumption on the $q$-$m_{\rm NS}$ correlation, so that the expected chirp mass function is no longer a simple convolution of the neutron star mass function with itself.
    That is also one reason why NSBH is favoured at the current stage, where the dependence on $q$ is solely addressed by the GW waveform, without exploiting any unclear presumption. 
    Furthermore, the measurement precision for $q$ in NSBH merger is better than is in BNS merger, thanks to the long-lasting waveform during NSBH merger event.
    However, the synergy of BNS and NSBH to deliver a joint constraint on $H_0$ is still foreseen in the future upon the rapid accumulation of both EM and GW data.
    
	\section{Conclusions} \label{sect: conclusion}
	
	We have demonstrated here that NSBH events provide an independent and potentially competitive means of estimating $H_0$ by relying on the characteristic Chandrasekhar mass of neutron stars - using only the GW data, and without the need for a supporting EM redshift. 
    This is thanks to the inherently narrow range of neutron star masses measured to lie in a narrow mass range 
    $m_{\rm NS} \approx 1.4 \pm 0.1 M_\odot$,
    provides a simple physically well understood `standard siren' for interpreting the GW waveforms of the GW events of this class. 
    At least one NSBH event is now undisputed, namely GW200115 with a redshifted NS mass of $1.48 M_\odot$ determined in the observer frame from the LVK waveform analysis and with a distances estimated to be $283^{+62.1}_{-82.7}$ Mpc using the measured strain amplitude. 
    Reconsideration of other GW events recorded prior to GW200115 may now be redefined with an NSBH classification, namely event GW190426 and possibly also GW190917 (for which a less conservative `prior' on the rate of NSBH merging is forced by the clear detection of GW200115). 
    Nonetheless we have shown that it is possible to obtain a consistent constraint in $H_0$ from the existing NSBH events, albeit with larger distance uncertainty due to relatively low SNR in the detections.
	
	We may look forward to upgraded sensitivity that would allow NSBH events with higher SNR, i.e. better defined waveforms for which the degeneracy with orbital/spin parameters can be much improved allowing significantly more precise distance estimation and hence better defined $H_0$ that will be defined from a joint analysis of a larger sample of future NSBH events. 
    We have emphasised that even with current sensitivity, with only 10 more useful NSBH events we can anticipate $\simeq$ 20\%. 
	
	We have emphasised that this NSBH based method for determining $H_0$ is unlike that of BNS events where an external redshift is required for a precision measurement of $H_0$, and for which only one BNS case has been detected this way (GW170817) and rather fortuitously via gamma-ray flare time coincidence and the unusually close proximity of this BNS event (GW170817). 
    For NSBH we have the advantage of a much larger detection horizon thanks to the presence of the associated massive black hole that enhances the chirp mass and hence the detectable SNR. Of course for NSBH events with associated EM emission whereby an independent redshift can be established, the precision on $H_0$ is much improved. Thus it should be regarded as a priority to be prepared for prompt followup of NSBH candidate GW events, but nevertheless this is not a requirement of our method, for which GW alone is sufficient for competitive precision on $H_0$ at current sensitivities. 
	
	\appendix
	
	



\bibliographystyle{mnras}
\bibliography{ref} 




\appendix

\section{Remarks on LVK Likelihood Evaluation} \label{app: lvk-likelihood-remark}
We note that the \textsc{pycbc} and \textsc{Bilby} GW parameters inference package used the accelerated method outlined in \citep{gw-param-accelerate} to compare between model templates and the data, and thus evaluating the likelihood. The method relies on the factorization of extrinsic parameters (in particular, $d_L$) and intrinsic parameters ($m_c, \eta$). As the effect of extrinsic parameters are shiftings and scalings of the waveform, in contrary to the expensive differential equation solvers involved for instrinsic parameters, analytical marginalization over extrinsic parameters are allowed. 

For almost the same reason, the aforementioned data pipeline does not take the redshift as input, templates evaluated at any distance are not redshift corrected, thus the corresponding output intrinsic parameters $\mathcal{M}_c$ are instead the redshift-uncorrected parameters. The MCMC samples provided by LVK determine only the distance $d_L$, the redshift samples are not used in the waveform fitting at all, and is obtained via trivially applying the assumed P18 cosmology with $H_0 = 67.9$ km s$^{-1}$ Mpc$^{-1}$. 


\bsp	
\label{lastpage}
\end{document}